\documentclass[preprint]{imsart}

\usepackage{amsthm,amsmath,amssymb}
\usepackage{subcaption}
\usepackage[ruled,vlined]{algorithm2e}
\usepackage{tabularx}
\usepackage{float}
\usepackage{fancybox}
\usepackage{bbold}
\RequirePackage[T1]{fontenc}
\usepackage{graphicx}
\RequirePackage[colorlinks,citecolor=blue,urlcolor=blue]{hyperref}

\usepackage[shortlabels]{enumitem}

\makeatletter
\def\namedlabel#1#2{\begingroup
	#2%
	\def\@currentlabel{#2}%
	\phantomsection\label{#1}\endgroup
}
\makeatother

\startlocaldefs
\endlocaldefs
\newtheorem{theorem}{Theorem}

\newtheorem{lemma}[theorem]{Lemma}
\newtheorem*{definition}{Definition}
\newtheorem{proposition}[theorem]{Proposition}
\newtheorem*{definitionP-Gap}{Definition (Pseudo-spectral gap)}

\newcommand\PG{\mathop{\mbox{$\rm P$-$\rm Gap$}}}
\newcommand{\G}{\mathrm{Gap}}
\newcommand{\GV}{\G \(\frac{1}{s}\)}
 \renewcommand{\epsilon}{\varepsilon}
 
 \renewcommand{\(}{\left(}
 \renewcommand{\)}{\right)}
 
  \newcommand{\po}{p^{\mathrm{opt}}}
 \newcommand{\argmax}{\rm argmax}
  
 \newcommand{\lm}{\lambda_{min}}
  \renewcommand{\d}{\rm diag}

 \newcommand{\diag}{\rm diag}
 \newcommand{\lmx}{\lambda_{max}}
  \newcommand{\<}{\left <}
 \renewcommand{\>}{\right >}
 \newcommand{\Sn}{\widehat{\Sigma}_n}
  \newcommand{\Qn}{\widehat{Q}_n}
 \newcommand{\Dn}{\widehat{D(p)}_n}
  
 \newcommand{\f}{f_n}
 \newcommand{\sas}{\sigma^2_{as}}
   \newcommand{\av}{\sigma^2_{as}}
  
  \newcommand{\wt}{w^{\rm new}}
  
  \newcommand{\Qt}{{Q^{\rm ext}}}
  \newcommand{\St}{{\Sigma^{\rm ext}}}
  \newcommand{\Qnt}{{Q^{\rm ext}_n}}
  \newcommand{\Snt}{{\Sigma^{\rm ext}_n}}
  \newcommand{\Dt}{{D^{\rm ext}_w}}
  \newcommand{\Dnt}{{D^{\rm ext}_n(w)}}
   
\newcommand{\E}{\mathbb{E}}
  
\makeatletter
\newcommand{\customlabel}[2]{%
	\protected@write \@auxout {}{\string \newlabel {#1}{{#2}{\thepage}{#2}{#1}{}} }%
	\hypertarget{#1}{#2}
}
\makeatother

\allowdisplaybreaks

\begin{document}

\begin{frontmatter}

\title{ADAPTING THE GIBBS SAMPLER}
\runtitle{Adapting The Gibbs Samplers}


\begin{aug}
\author{\fnms{Cyril} \snm{Chimisov}\thanksref{}\ead[label=e1]{K.Chimisov@warwick.ac.uk}},
\author{\fnms{Krzysztof} \snm{{\L}atuszy{\'n}ski}\thanksref{}\ead[label=e2]{K.G.Latuszynski@warwick.ac.uk}}
\and
\author{\fnms{Gareth O.} \snm{Roberts}\thanksref{}
\ead[label=e3]{Gareth.O.Roberts@warwick.ac.uk }}


\runauthor{C. Chimisov et al.}

\affiliation{University of Warwick\thanksmark{m1}}

\address{Department of Statistics\\
University of Warwick\\
Coventry\\
CV4 7AL\\
United Kingdom\\
\printead{e1}\\
\phantom{E-mail:\ }\printead*{e2}\\
\phantom{E-mail:\ }\printead*{e3}}

\end{aug}

\begin{abstract}

The popularity of Adaptive MCMC has been fueled on the one hand by its success in applications, and on the other hand, by mathematically appealing and computationally straightforward optimisation criteria for the Metropolis algorithm acceptance rate (and, equivalently, proposal scale). Similarly principled and operational criteria for optimising the selection probabilities of the Random Scan Gibbs Sampler have not been devised to date.

In the present work we close this gap and develop a general purpose Adaptive Random Scan Gibbs Sampler that adapts the selection probabilities. The adaptation is guided by optimising the $L_2-$spectral gap for the target's Gaussian analogue \cite{Amit1996, Roberts1997}, gradually, as target's global covariance is learned by the sampler. The additional computational cost of the adaptation represents a small fraction of the total simulation effort.

We present a number of moderately- and high-dimensional examples, including truncated Gaussians, Bayesian Hierarchical Models and Hidden Markov Models, where significant computational gains are empirically observed for both, Adaptive Gibbs, and Adaptive Metropolis within Adaptive Gibbs version of the algorithm. We argue that Adaptive Random Scan Gibbs Samplers can be routinely implemented and substantial computational gains will be observed across many typical Gibbs sampling problems.

We shall give conditions under which ergodicity of the adaptive algorithms can be established.
\end{abstract}



\end{frontmatter}


\tableofcontents

\section{Introduction} \label{section introduction}

Markov Chain Monte Carlo (MCMC) methods is a powerful tool to estimate integrals $\int_{\mathcal{X}} f(x) \pi( {\rm d} x)$ of some function $f$ with respect to (w.r.t.) some probability measure $\pi$ on a measurable space $\mathcal{X}$.

The idea behind the MCMC technique is fairly simple. First, we need to construct a Markov kernel $P$ that has $\pi$ as its stationary distribution, i.e., $\int_\mathcal{X} P(x, \cdot) \pi( {\rm d} x) = \pi(A)$. Then, we run a Markov chain using the kernel $P$ to obtain samples $\{X_i\}_{i=1}^n$, which can be used to estimate $\int f {\rm d} \pi$ by the average $\frac{1}{n}\sum_{i=1}^n f(X_i)$ (see, e.g., \cite{Liu2008}).

While designing the kernel  $P$ is easy (e.g., one can come up with dozens of proposals in the  Random  Walk Metropolis (RWM) scheme), identifying kernels $P$ for which  $\frac{1}{n}\sum_{i=1}^n f(X_i)$ does not converge excessively slowly is a hard problem. 

Typically, the user has to choose a kernel from a parametrised family $P_\gamma$, $\gamma\in \Gamma$  with a common target stationary distribution $\pi$. For example, $\Gamma$ may represent a collection of proposals for the RWM algorithm or a set of selection probabilities (weights) for the Random Scan Gibbs Sampler (RSGS) that are used to decide which coordinate to update next.

A naive approach to find a good parameter $\gamma$ would require the user to re-run the MCMC algorithm many times before a good Markov kernel candidate $P_\gamma$ is found.  

An alternative idea is to come up with an adaptation rule which changes the value of $\gamma$ during the run of the Markov chain, gradually, as further information is acquired by the chain. This approach is known as  Adaptive  MCMC (AMCMC) algorithms and is very attractive in practice since it frees users from the cumbersome process of hand-tuning the parameters and potentially accelerates convergence to the target distribution. Formally, an  AMCMC algorithm produces a chain $X_n$ by repeating the following two steps.
\begin{enumerate}[label*=(\arabic*)]
	\item \label{alg:amcmc:sample} Sample $X_{n+1}$ from $P_{\gamma_{n}} \(X_{n}, \cdot \)$;
	\item \label{alg:amcmc:update} Given $\{X_0,.. , X_{n+1}, \gamma_0,.., \gamma_n\}$ update $\gamma_{n+1}$ according to some adaptation rule.
\end{enumerate}
After running an adaptive chain, we can use its output in the same way as if it were a usual MCMC chain in order to estimate $\int f {\rm d} \pi$. Note that the adaptive chain is not Markov in general, making its analysis particularly complicated.

First guidance on how to construct an ergodic  AMCMC is proposed by Gilks et al.  \cite{Gilks1998a}, where the authors allow any kind of adaptations to take place but only at the regeneration times of the underlying Markov chains. Unfortunately, the algorithm is inefficient in high dimensional settings since the regeneration rate deteriorates exponentially in dimension. More practical conditions are due to Roberts {\&} Rosenthal \cite{Roberts2007} and are known as diminishing and containment conditions \ref{cond:diminishing}, \ref{cond:containment}, which we discuss in Section \ref{section Ergodicity}.

Even though one has theoretical results that help establish convergence of the adaptive algorithms, no less important challenge is to come up with an adaptation scheme for the Step \ref{alg:amcmc:update} of the AMCMC algorithms. For the RWM, the adaptation rule is based on the approximation of the optimal Gaussian proposal that has been studied by \cite{Gelman1996, Roberts1997c, Roberts2001a, Bedard2007, Bedard2008}. The authors have noticed that in high dimensional spaces, the RWM behaves like a diffusion, so that one has to optimise the proposal variance in order to maximise convergence speed of the limiting diffusion. In dimensions $d\geq5$, the optimal covariance matrix for the Gaussian proposal according to \cite{Roberts1997c, Roberts2001a} is $\alpha \Sigma$, where $\Sigma$ is a $d\times d$ covariance matrix of the target distribution and the scaling parameter $\alpha > 0$ is chosen so that the average acceptance ratio of the algorithm is $0.234$. In practice both $\Sigma$ and $\alpha$ are not known in advance but can be learned in the Step \ref{alg:amcmc:update} of the AMCMC algorithm. Successful adaptive RWM algorithms have been proposed by \cite{Haario2001, Roberts2009, Vihola2012}.    Convergence properties of the Adaptive RWM have been extensively studied in the literature, e.g., \cite{Andrieu2007a, Atchade2010, Atchade2005, Saksman2010, Vihola2011, Vihola2012}.

At the same time, the Random Scan Gibbs Sampler (RSGS) and  Metropo-{\newline}lis-within-Gibbs (MwG) algorithms are very popular in practice.  Recall that the RSGS  at every iteration chooses a coordinate $i$ with probability $p_i$ and updates it from its full conditional distribution. If the full conditional distribution is expensive or impossible to sample from, then a proposal is generated for the direction $i$ from some proposal distribution $Q_i$, followed by the Metropolis-Hastings  acceptance/rejection procedure. The corresponding algorithm  is called Metropolis-within-Gibbs.  
  
Usually, uniform selection probabilities $p_i$ are used, while we argue that this is often a sub-optimal strategy. To date, there is no guidance on the optimal choice of the selection probabilities as noticed in \cite{Latuszynski2013}. 

A possible solution is to use those probabilities that maximise the $L_2-$spectral gap (hereafter, spectral gap) of the corresponding algorithm. Of course, estimating the spectral gap is a challenging problem. On the other hand, if the target distribution is normal, then for the RSGS there is an explicit formula (\ref{spectral gap for fixed p}) for the spectral gap.  Since  (\ref{spectral gap for fixed p}) depends only on the correlation structure of the target distribution, the equation (\ref{spectral gap for fixed p}) may be optimised  for an arbitrary target distribution resulting in some selection probabilities $\po$, that we call {\it pseudo-optimal}. The corresponding value of (\ref{spectral gap for fixed p}) at $\po$ is the {\it pseudo-spectral gap}.
 
In Bayesian Analysis, by virtue of Bernstein-von Mises Theorem (see Section 10.2 of \cite{Vaart2000}), under certain conditions, given sufficient amount of observations, the posterior distribution is well approximated by an appropriate Gaussian. Thus if one applies the RSGS to sample from the posterior, the pseudo-optimal weights $p_i$ might represent a good approximation to the true optimal weights that maximise the spectal gap. Interestingly, as we demonstrate by simulations in Section \ref{section simulations}, even if the target distribution is discrete, the pseudo-optimal weights might still be advantageous over the uniform selection probabilities. 
  
Since the pseudo-optimal selection probabilities are a function of the correlation structure of the target distribution, which is usually not known, and optimising the pseudo-spectral gap function (\ref{spectral gap for fixed p}) is a hard problem (see, e.g., \cite{Overton1988}), we develop a general purpose Adaptive Random Scan Gibbs Sampler (ARSGS) that adapts the selection probabilities on the fly.  

We also find that a special case of the MwG algorithm, namely, Random Walk Metropolis within Gibbs (RWMwG) algorithm, may be significantly improved by adapting both the proposal distribution (for instance, as suggested in \cite{Jeffrey2011}) and the  underlying selection probabilities in the same manner as for the RSGS.

Because the implementation of the adaptive algorithms is easy and the additional computational cost is often negligible compared to the total computational effort, we argue that the algorithms could be routinely implemented. We demonstrate in Section \ref{section simulations} that the ARSGS and ARWMwAG algorithms speed up convergence to the target distribution for many typical Gibbs sampling problems. 
 
Finally, we introduce a notion of {\it local simultaneous geometric drift} condition \ref{cond:local_simultaneous_drift} in Section \ref{section Ergodicity}. It turns out that for the RSGS it is a natural property to have as we demonstrate in Theorem \ref{theorem local simultaneous drift}. In Theorem \ref{theorem ARGS convergence} we prove convergence of the modified ARSGS under the local simultaneous geometric drift condition.

The paper is organised as follows. In Section \ref{section normal spectral gap} we exploit ideas of Amit  \cite{Amit1991a, Amit1996} and  Roberts \& Sahu \cite{Roberts1997} to derive the formula for the spectral gap for a particular case of sampling from the Multivariate Normal distribution using the RSGS. For a general target distribution, we introduce the concept of pseudo-spectral gap and pseudo-optimal selection probabilities in Section \ref{section pseudo-spectral gap} and  demonstrate potential advantage of the pseudo-optimal weights on toy examples studied in Section \ref{section example}. Derivation of the ARSGS and ARWMwAG algorithms is presented in Sections \ref{section Adaptive Gibbs} and \ref{section AMWMwAG} respectively. Convergence properties of the adaptive algorithms are discussed in Section \ref{section Ergodicity}.  We provide simulation study and discuss computational cost of the adaptive algorithms  in Section \ref{section simulations}. Unless stated otherwise, the proofs are presented in the Supplementary Material section \ref{SUPPLEMENTARY  MATERIAL}.

\section{RSGS spectral gap for Multivariate Gaussian distribution} \label{section normal spectral gap}

In this section we consider the RSGS for the normal target distribution and establish an explicit representation of the spectral gap in Theorem \ref{theorem spectral gap for normal distribution}.
One may skip all the technical details and notice only that the spectral gap in this case relies solely on the correlation structure of the target distribution and the selection probabilities.

Let $\pi$ be a distribution of interest in $\mathbb{R}^d$. Let  $\Sigma$ and $Q = \Sigma^{-1}$ denote the covariance matrix of $\pi$ and its inverse respectively, where we assume throughout the paper that $\Sigma$ is positive-definite. Partition $Q$ into blocks $Q=\(Q_{ij}\)_{i,j =1}^s$ where $Q_{ij}$ is a $r_i \times r_j$ matrix, $\sum_{i=1}^s r_i = d$. For vectors $x\in \mathbb{R}^d$ introduce splitting $x = (x_1,..,x_s)$, where $x_i$ is a vector in $\mathbb{R}^{r_i}$ so that $x_i = \(x_{i1},..,x_{ir_i}\)$.

Given a probability vector  $p=(p_1,..,p_s)$ (i.e., $p_i> 0,\ \sum_{i=1}^s p_i=1$),  RSGS($p$) is a Markov kernel that at every iteration chooses a subvector $x_i = \(x_{i1},..,x_{ir_i}\)$ with probability $p_i$ and updates it from the conditional distribution $\pi(x_i | x_{-i})$ of $x_i$ given  $x_{-i}:=(x_1,..,x_{i-1},x_{i+1},..,x_s)$.  In other words, the RSGS($p$) is a Markov chain with kernel
\begin{align}\label{kernel}
P_p (x,A) = \sum_{i=1}^s p_i  Pr_i(x, A),
\end{align}
where $A$ is a $\pi-$measurable set, $x\in \mathbb{R}^d$ and $Pr_i$ is a kernel that stands for updating $x_i$  from the full conditional distribution  $\pi(x_i | x_{-i})$. We call the kernel $Pr$ since it is in fact a projection operator (i.e., $Pr^2 = Pr$) acting on the set of 
the space of square integrable functions $L_2(\mathbb{R}^d, \pi)$ with respect to $\pi$. For $\pi-$integrable functions $f$, let $\pi(f):=\int f {\rm d} \pi$ and $\(P_p f\)(x) := \int f(y) P(x, {\rm d} y).$ 

\begin{definition}
	Let $\rho = \rho(p)>0$ be the minimum number such that for all $f\in L_2 (\mathbb{R}^d, \pi)$ and $r>\rho$,
	\begin{align}\label{def:spectral gap}
	\lim_{n\to \infty}  r^{-2n}\mathbf {E}_{\pi} [\{\(P_p^n f\) (x)-\pi(f)\}^2]=0.
	\end{align}
	Then $\rho$ is called the {\it ${L_2-}$rate of convergence} in $L_2 (\mathbb{R}^d, \pi)$ of the Markov chain with the kernel  $P_p$. The value $1- \rho$ is called the {\it $L_2-$spectral gap} (or simply {\it spectral gap}) of the kernel $P_p$.
\end{definition}

In the case when $s = d$ and the selection probabilities are uniform, i.e., $p = (\frac{1}{d},.., \frac{1}{d})$, Amit \cite{Amit1996} provides a formula for the spectral gap. Here we generalise Amit's result by essentially changing $p_i$ for $\frac{1}{s}$ in the proof of Theorem 1 in \cite{Amit1996}.

It is easy to see that the RSGS kernel is reversible w.r.t. the target distribution $\pi$. It is known that if the spectrum of kernel $P_p$ (considered as an operator on $L_2 (\mathbb{R}^d, \pi)$) consists of eigenvalues only, the $L_2-$rate of convergence is given by the second largest eigenvalue of the kernel $P_p$ (follows, e.g, from Theorem 2 and the following remark in  \cite{Roberts1997b}).

There are two key steps to establish an explicit formula for the rate of convergence $\rho (p)$.
\hfill\\

\noindent {\bf Step 1.} For the kernels $P_p$, find finite dimensional invariant subspaces $S_k$ (i.e., $P_p S_k \subset S_k$) in $L_2(\mathbb{R}^d,\pi)$ by considering action of $P_p$ on the orthonormal basis of Hermite polynomials.

\noindent{\bf Step 2.} Identify the subspace $S_k$ with the maximum eigenvalue less than one. 

To clarify the steps we need to introduce some additional notations. Without loss of generality, suppose that $\pi$ has zero mean. 

Let $K = \sqrt{Q}$  be the symmetric square root of $Q$ defined through the spectral decomposition , i.e., if for an orthogonal matrix $U$ (i.e., $U^{\mathsf{T}}U = I$), $Q=U\d (\lambda_1,..,\lambda_n) U^{\mathrm{T}}$, then 
$K = U\d (\sqrt{\lambda_1},..,\sqrt{\lambda_n})U^{\mathrm{T}}$. Set 

\begin{align}\label{matrix_D_i}
	D_i = {\rm diag}(0,..,Q^{-1}_{ii},..,0),
\end{align}
where we stress that $D_i$ is a $d\times d$ matrix with $Q^{-1}_{ii}$ being at the same place as in partition $(Q^{-1}_{ii})_{i=1}^s$.

For $\alpha=(\alpha_1,..,\alpha_d)\in Z^d_+$ let $\alpha! = \alpha_1 !.. \alpha_d !$, $|\alpha|=\alpha_1+..+\alpha_d$.  Define $h_k$ to be the {\it Hermite polynomial} of order $k$, i.e.,

$$h_k(x) = (-1)^k \exp\(\frac{x^2}{2}\) \frac{d }{dx^k} \exp\(-\frac{x^2}{2}\), \ x \in \mathbb{R}^d.$$

Set  $H_\alpha (x)=\frac{1}{\sqrt{\alpha!}} h_{\alpha_1}(x_1).. h_{\alpha_d} (x_d)$, $H_0(x): = 1$. The next lemma summarizes Steps   1 and 2 above.

\begin{lemma}\label{lemma hermite polynomials}
	$\left\{{H_\alpha (K x)} |\ \alpha \in Z^n_+\right\}$ form an orthonormal basis in $L_2(\mathbb{R}^d,\pi)$ and  for all integers $k\geq 0$, spaces 
	$$S_k := {\rm span} \Big\{H_\alpha (Kx) \Big| \  |\alpha|=k\Big\},$$
	spanned by $\{H_\alpha (Kx) |\  |\alpha|=k\}$, are finite dimensional and  $P_p - $invariant (i.e., $P_p (f) \in S_k$ for all $f\in S_k$). Moreover, for all $k\geq 0$,
	$$\lmx(P_p|_{S_1})\geq\lmx(P_p|_{S_{k}}),$$
	where $\lmx(\cdot)$ is the maximum eigenvalue and $P_p|_{S_k}$ is a restriction of $P_p$ on $S_k$.
\end{lemma} 

Lemma \ref{lemma hermite polynomials} immediately implies that $\G (p) = 1 - \lmx (P_p|_{S_1})$ and the next theorem provides a representation of $\G (p)$ through  the correlation structure of the target distribution.

\begin{theorem} \label{theorem spectral gap for normal distribution}
	The $L_2-$spectral gap in the RSGS(p) scheme for the Gaussian target  distribution with precision matrix $Q$ is given by 
	\begin{align}\label{spectral gap for fixed p}
	\G(p) =1 - \lmx(F_1),
	\end{align}
	where 
	\begin{align}\label{Amit result}
	F_1  =I - K\( \sum_{i=1}^s p_i  D_i   \)K,
	\end{align}
	$D_i$ is given by (\ref{matrix_D_i}), and $K = \sqrt{Q}$.
\end{theorem}

Since $S_1$ is a set of linear functions, Lemma \ref{lemma hermite polynomials} also implies

\begin{theorem} \label{theorem 2nd eigenfunction}
	Consider a Gibbs kernel $P_p$ that corresponds to a normal target distribution $\pi$. Then the second largest eigenfunction of $P_p$ in $L_2(\mathbb{R}^d, \pi)$  is a linear function.
\end{theorem}

We end this section by comparing formula (\ref{spectral gap for fixed p}) with the results by Roberts \& Sahu \cite{Roberts1997}. Consider the case when $p_1=..=p_s=\frac{1}{s}$ and introduce a matrix 

\begin{align}\label{DUGS matrix}
A=I - {\rm diag}(Q^{-1}_{11},..,Q^{-1}_{ss}) Q,
\end{align}

The following lemma will be useful throughout the paper and can be easily obtained.
\begin{lemma} \label{lemma isometry argument}
	Let $A$ and $B$ be two $d\times d$ matrices. Then $AB$ and $BA$ have the same eigenvalues.
\end{lemma}
Lemma \ref{lemma isometry argument} implies that the spectrum (the set of all eigenvalues) of $A$ defined in (\ref{DUGS matrix}) is equal to the spectrum of
$$I- K {\rm diag}(Q^{-1}_{11},..,Q^{-1}_{ss}) K.$$ 

One can easily see that $T^{(i)}:=I-KD_i   K$ is a projection matrix, hence
\begin{align*}
&I- K {\rm diag}(Q^{-1}_{11},..,Q^{-1}_{ss}) K = I+\sum_{i=1}^sT^{(i)} - sI\geq (1-s) I,
\end{align*}
and the minimum eigenvalue of $A$ is bounded below by $(1-s)$.  Therefore, (\ref{spectral gap for fixed p}) is equivalent to

$$\G \(\frac{1}{s}\) = \lmx \(\frac{1}{s}\( (s-1)I+A\)\) = \frac{1}{s}\Big( s-1+\lmx(A)\Big),$$ 
where $\G \(\frac{1}{s}\)$ is the spectral gap of the RSGS with the uniform selection probabilities.

The last equation is the representation of the spectral gap in Theorem 2 of \cite{Roberts1997}.

\section{Pseudo-spectral gap} \label{section pseudo-spectral gap}

For a general target distribution computing the spectral gap is not feasible. But one can always deal with its normal counterpart (\ref{spectral gap for fixed p}) which we call {\it pseudo-spectral gap}. Optimizing  (\ref{spectral gap for fixed p}) over all possible selection probabilities $p$ leads to the notion of {\it pseudo-optimal selection probabilities}.

As mentioned in Section \ref{section introduction}, in many Bayesian settings  Bernstein-von Mises theorem (see, e.g, Section 10.2 of \cite{Vaart2000}) applies, that is,  under certain conditions the posterior distribution converges to normal in the total variation norm.  Thus we hope that the pseudo-spectral gap of RSGS is a meaningful approximation to the true value of the spectral gap and the pseudo-optimal weights are close to the ones that maximise the spectral gap.

In fact, as we will see in Section \ref{section simulations}, where we sample from the Truncated Multivariate Normal distribution and the posterior in Markov Switching Model, if the correlation matrix is well-informative about the dependency structure of the target distribution, running the RSGS with the pseudo-optimal weights instead of the uniform ones, may substantially fasten the convergence, even if the target distribution has discrete components. 

To formally define the pseudo-spectral gap, we need a couple of additional notations.
$$ {\Delta}_{s-1}:=\{\bar{p}\in \mathbb{R}^{s-1} | \bar{p}_i>0,\ i=1,..,{s-1};\ 1-\bar{p}_1-..-\bar{p}_{s-1}>0\}$$
is a convex set in $\mathbb{R}^{s-1}$, so that $\Delta_{s-1}$ defines a set of $s-$dimensional probability vectors $p=(p_1,..,p_s)$ and we write $p\in \Delta_{s-1}$ meaning $(p_1,..,p_{s-1}) \in \Delta_{s-1}$. 

Let $\lm(\cdot)$ and $\lmx(\cdot)$ denote the minimum and the maximum eigenvalues of a matrix respectively. As before, for a covariance matrix $\Sigma$, $Q = \Sigma^{-1}$, $K  = \sqrt{Q}$. For probability weights $p=(p_1,..,p_s)$, let
\begin{align}\label{matrix_D_p}
	D_p=\diag(p_1 Q_{11}^{-1},...,p_s Q_{ss}^{-1})
\end{align}
be a $d \times d$ block-diagonal matrix.

\begin{definitionP-Gap}
	For arbitrary distribution $\pi$ with precision matrix $Q$, and any probability vector $p\in \Delta_{s-1}$, the { pseudo-spectral gap} for RSGS(p) is defined as
	\begin{align} \label{pseudo-spectral gap}
	\PG(p): = 1 - \lmx \(I - K D_p K\),
	\end{align}
	which due to Lemma \ref{lemma isometry argument} can be written as
	\begin{align} \label{pseudo-spectral gap 2}
	\PG(p) = 1- \lmx \(I -   D_p  Q\)=\lm \(   D_p  Q\).
	\end{align}
	
	Weights  $\po=\(\po_1,..,\po_s\) \in \Delta_{s-1}$ are called { pseudo-optimal} for RSGS if they maximize the corresponding pseudo-spectral gap, i.e,
	\begin{align} \label{definition pseudo-optimal}
	\po=\underset{{p\in \Delta_{s-1}}}{\argmax} \lm \(   D_p  Q\).
	\end{align}
\end{definitionP-Gap}

\noindent {\bf Remark.} It follows from Section \ref{section normal spectral gap}, that for RSGS(p) the pseudo-spectral and the spectral gap are the same if the target distribution is normal.  \\

Useful observation for both theoretical and practical purposes is the uniqueness of the pseudo-optimal weights.
\begin{theorem}\label{theorem uniqueness}
	There exists a unique solution for  (\ref{definition pseudo-optimal}).
	
\end{theorem}

We conclude this section by presenting an upper bound on the possible improvement of the spectral gap of RSGS($\po$) compared to the spectral gap of the vanilla chain, i.e., the chain with uniform selection probabilities.

\begin{theorem}\label{theorem upper bound} Let $\G (p)$ be the spectral gap of RSGS($p$) and $\GV$ be the spectral gap of the vanilla chain, i.e., the RSGS with uniform selection probabilities. Then for any probability vectors $p$ and $q$
	$$\G(p)\leq \(\max_{i = 1,..,s}  { \frac{p_i}{q_i}}\) \G(q),$$
	in particular,
	\begin{align} \label{gap upper bound}
	\G(p)\leq \(\max_{i = 1,..,s}  { s p_i}\) \GV,
	\end{align}
	where $s$ is the number of components in the Gibbs sampling scheme.
\end{theorem}

\noindent {\bf Remark.} Theorem \ref{theorem upper bound} implies 

\begin{align} \label{pseudo-gap upper bound}
\PG(p)\leq \(\max_{i = 1,..,s}  { s p_i}\) \PG\(\frac{1}{s}\),
\end{align}
where $\PG(\frac{1}{s})$ is the pseudo-spectral gap for the vanilla chain.

Theorem \ref{theorem upper bound} states that the maximum gain one can get by using non-uniform selection probabilities is bounded by $s$ times - the number of blocks in the Gibbs sampling scheme. Thus we expect the pseudo-optimal weights to be particularly useful in high dimensional settings.

\section{Motivating examples} \label{section example}

The pseudo-optimal weights (\ref{definition pseudo-optimal}) have complicated interpretation as we will see in the following examples.
\\

\noindent {\bf Example \customlabel{example1}{1} \hspace{-4mm}.}
In  case where the correlation matrix of the target distribution has blocks of highly correlated coordinates, one would  prefer to update them more frequently than the others. In this section we construct an artificial example where the upper bound in (\ref{pseudo-gap upper bound}) is $\frac{d}{2} \G\(\frac{1}{d}\)$.  Consider a target distribution in $\mathbb{R}^{d }$, $d = 2 k$ with correlation and normalised precision (inverse covariance) matrices given respectively by their block form, i.e., ${\rm Corr} = \(C_{ij}\)_{i,j = 1}^k$, $Q = \(Q_{ij}\)_{i,j=1}^k$, where $C_{ij}$ and $Q_{ij}$ are $2\times 2$ matrices such that all $Q_{ij}$, $C_{ij}$ are zero matrices if $i\neq j$ and for all $i = 1,..,k$

\begin{align*}
C_{ii}= \left( \begin{array}{cc} 1 & -\rho_i \\
            -\rho_i & 1 
             \end{array} \right),\ \ \ Q_{ii}= \left( \begin{array}{cc} 1 & \rho_i \\
            \rho_i & 1 
             \end{array} \right),
\end{align*}
where we assume $\rho_i\geq 0$, $i=1,..,k$. Assume one wants to apply the coordinate-wise RSGS to sample from a distribution with the above correlation matrix.

\begin{proposition}\label{proposition example 2k*2k}
Let the inverse covariance matrix $Q$ be as above. Define 
\begin{align}\label{alpha k}
\alpha_i = \frac{\prod_{l=1, l\neq i}^k (1-\rho_l)}{\sum_{l=1}^k\prod_{j=1,j\neq l}^k (1-\rho_j)}.
\end{align}
Then the pseudo-optimal weights are given by
\begin{align} \label{expression optimal k}
\po_{2i-1}=\po_{2i}=\frac{\alpha_i}{2}.
\end{align}
The corresponding $\PG$ is 
\begin{align}\label{optimal rate of convergence example}
\PG \(\po\) = \frac{\prod_{l=1}^k (1-\rho_l)}{2 \sum_{l=1}^k\prod_{j=1,j\neq l}^k (1-\rho_j)}.
\end{align}
\end{proposition}

Without loss of generality assume $\rho_1=\max\{\rho_1,..., \rho_k\}$. We shall compare pseudo-spectral gaps of the vanilla chain with RSGS($\po$). One can easily  obtain that  the pseudo-spectral gap of the vanilla chain is given by

$$\PG \(\frac{1}{d}\) = \frac{1}{d} (1-\rho_1).$$
Simple calculations yield
\begin{align*}
&\lim_{\rho_1\to 1}\frac{\PG \(\frac{1}{d}\)}{\PG \(\po\)}=\lim_{\rho_1\to 1}\frac{1-\rho_1}{2k \(\frac{\prod_{l=1}^k (1-\rho_l)}{ 2 \sum_{l=1}^k\prod_{j=1,j\neq l}^k (1-\rho_j)}\)}=\\
&=\lim_{\rho_1\to 1}\frac{1}{k} \frac{\(\sum_{l=1}^k\prod_{j=1,j\neq l}^k (1-\rho_j)\)}{   \prod_{l=2}^{k} (1-\rho_l)}=\frac{1}{k} = \frac{2}{d}.
\end{align*}
Moreover, 
$$\lim_{\rho_1\to 1} \( \max_i d \po_i  \)=\frac{1}{k} = \frac{2}{d}.$$

Thus we obtained a sequence of precision  matrices for which the pseudo-optimal weights improve the pseudo-spectral gap by $\frac{d}{2}$ times in the limit which is the upper bound in (\ref{pseudo-gap upper bound}). Notice, if the underlying target distribution is normal, the upper bound in (\ref{gap upper bound}) for the  $L_2-$spectral gap is approximated.

\noindent {\bf Remark.}  Corollary 1 to Theorem 5 of \cite{Roberts1997} implies that the spectral gap of Deterministic Update Gibbs Sampler (denoted by $\G \({\rm DUGS}\)$) for the normal target with a 3-diagonal precision $Q$ is greater  than the gap of the vanilla RSGS (i.e., with the uniform selection probabilities). Moreover, from Corollary 2 to Theorem 5 of \cite{Roberts1997}, $\underset{\rho_1\to 1}{\lim} \frac{\G \({\rm DUGS}\)}{\PG \(\frac{1}{d}\)} = 2$.  We constructed an example of a 3-diagonal precision matrix, where in dimensions greater than $6$, RSGS with  pseudo-optimal weights $\po$ converges $\frac{d}{4}$ times faster than DUGS for $\rho_1\to 1$.
\\

\noindent {\bf Example \customlabel{example2}{2}\hspace{-1mm}.} One mistakenly might conclude that significant gain from using the pseudo-optimal weights is achieved only if some of the off-diagonal entries of the covariance matrix are close to one . Here we provide a somewhat counter-intuitive example that demonstrates fallacy of such statement.

Consider a correlation matrix matrix given by $\Sigma^{(2)} = \(C_{ij}\)_{i,j=1}^d$, where $C_{ii} = 1$ for $i=1,..,d$, $C_{1i} =C_{i1}: =c_i \geq0 $ for $i= 2,..,d$ and all other entries $C_{ij} = 0$.

One can easily work out that the smallest eigenvalue of $\Sigma^{(2)}$, $\lm = 1-\sqrt{\sum_{i=2}^d c^2_i}$. Thus if $\lm>0$, then $\Sigma^{(2)}$ is a valid correlation matrix. Set $d=50$ and $c_i = \frac{1}{7.01} \approx 0.143$ for $i=2,.., 50$. 

We run the subgradient optimisation algorithm presented in Section \ref{section Adaptive Gibbs} in order to estimate $\po$. We estimate $\po_1\approx 0.484$, $\po_i \approx 0.01$ for $i =2, .. ,50$.  From (\ref{pseudo-spectral gap 2})
 the pseudo-spectral gap  is roughly $\frac{1}{1496}$, whilst $\PG \(\frac{1}{50}\)$ is roughly $\frac{1}{18294}$.
Thus if the target distribution is normal, the spectral-gap of the vanilla RSGS is improved by more than $12$ times. Note, however, all off-diagonal correlations are less than $0.143$.

\section{Adapting the Gibbs Sampler} \label{section Adaptive Gibbs}

In this section we derive the Adaptive Random Scan Gibbs Sampler (ARSGS) Algorithm \ref{alg:ARSGS}. We provide all the steps and intuition leading  towards the final working version of the algorithm presented in the end of the section.

The goal is to compute the pseudo-optimal weights (\ref{definition pseudo-optimal}) for the RSGS (\ref{kernel}). However, in practice the correlation matrix of the target distribution is usually not known. Thus we could proceed in the adaptive way, similarly to Haario et al. \cite{Haario2001}. Given output of the chain of length $n$, let $\Sn$, $\Qn$, and $\Dn$ be estimators of $\Sigma$, $Q$, and $D_p$ respectively built upon the chain output. For instance, one may choose the naive estimator
\begin{align}\label{naive estimator}
\Sn = \frac{1}{n}\(\sum_{i=0}^{n} X_i X_i^\mathrm{T} - (n+1) \overline{X}_{n} \overline{X}_{n}^\mathrm{T}\),
\end{align}
where $X_n$ is the chain output at time $n$ and $\overline{X}_{n}$ is a sample mean of the output up to time $n$.

\begin{algorithm}
	\caption{Adaptive Random Scan Gibbs Sampler (general idea)} \label{alg:AGS general idea}
	Generate a starting location $X_0\in \mathbb{R}^d$. Set an initial value of $p^0 \in \Delta_{s-1}$.  Choose a sequence of positive integers $\(k_m\)_{m=0}^{\infty}$. Set $n=0$, $i=0$.\\
	{\bf Beginning of the loop}
	\begin{enumerate}[label={\arabic*}.,ref={\arabic*}]
		\item $n := n+k_i$. Run RSGS($p^{i}$) for $k_{i}$ steps;
		\item Re-estimate $\Sn$ and $\Dn$;
		\item \label{alg:AGS general idea:adapt} Compute $p^{i+1}=\argmax_{p\in \Delta_{s-1}}\lm \(   \Dn  \Qn\)$;
		\item $i:=i+1$. 
	\end{enumerate}
	Go to {\bf Beginning of the loop} 
\end{algorithm}
\vspace{5mm}

The Algorithm \ref{alg:AGS general idea} summarises the above ideas. The algorithm is  limited by Step \ref{alg:AGS general idea:adapt}, where one needs to maximize the minimum eigenvalue. Maximising the minimum eigenvalue is known to be a complicated optimisation problem. There is vast literature covering optimisation problem in Step \ref{alg:AGS general idea:adapt} and we refer to \cite{Overton1988, Overton1992}, \cite{Chu1990}, and references therein. Unfortunately, the existing optimisation algorithms require computation of  the minimum eigenvalues of  $\lm \(   \Dn  \Qn\)$ which is not a reasonable way to waste computational resources  since we do not know the covariance matrix $\Sigma$ anyway. Therefore, we develop a new algorithm based on the subgradient method for convex functions (see Chapter 8 of \cite{Bertsekas2003}) applied to (\ref{definition pseudo-optimal}).

For $\epsilon>0$, introduce a contraction set of ${\Delta}_{s}$:
\begin{align}\label{def:contracted simplex}
{\Delta}^\epsilon_{s} : =\{ w \in \mathbb{R}^{s} | w_i \geq \epsilon,\ i=1,..,{s-1};\ 1-w_1-..-w_{s}\geq \epsilon\},
\end{align}
and consider $(d+1)\times (d+1)$ matrices 

\begin{align*}
\begin{aligned}[c]
&\Qt = {\rm diag} \( Q , 1\),\\
&\Snt = {\rm diag} \(\Sn, 1\),\\
&\Dt = {\rm diag}\( D_w,\ 1 - \sum_{i=1}^s w_i\),
\end{aligned}
\qquad
\begin{aligned}[c]
&\St = {\rm diag} \(\Sigma , 1\),\\
&\Qnt = {\rm diag} \(\Qn. 1\),\\
&\Dnt = {\rm diag} \( \widehat{D(w)}_n,\ 1 - \sum_{i=1}^s w_i\).
\end{aligned}
\end{align*}

Let us denote the target function 

\begin{align}\label{target function}
f(w)=\lm\(\Dt \Qt\) = \lm \(\sqrt{\Qt} \Dt \sqrt{\Qt}\),
\end{align}
where the last equality holds in view of Lemma {\ref{lemma isometry argument}}.  

Using the definition of the pseudo-optimal selection probabilities (\ref{definition pseudo-optimal}), one can easily verify the following proposition
\begin{proposition}\label{proposition extended pseudo-optimal points}
	The pseudo-optimal weights (\ref{definition pseudo-optimal}) can be obtained as a normalised solution of
	\begin{align*}
	w^{\star} = \underset{{w\in \Delta_{s}}}{\argmax} f(w),
	\end{align*}
	i.e.,
	\begin{align}\label{extended pseudo-optimal points}
	\po_j = \frac{w^{\star}_j}{w^{\star}_1+.. + w^{\star}_s},\ j=1,..,s,
	\end{align}
	Moreover, 
	$$\PG(\po) = \frac{1}{(w^\star_1+..+w^\star_s)} f(w^{\star}),$$
	where $f$ is defined in (\ref{target function}).
\end{proposition}

\noindent {\bf Remark.} One could easily avoid introducing the extended matrices $\St, \Qt$ by simply setting $p_s = 1 - p_1 - .. -p_{s-1}$ and treating function $\lm \(   \Dn  \Qn\)$ as a function of $s-1$ variables. However, we found empiricallym that such approach can significantly slow down convergence of the ARSGS Algorithm \ref{alg:ARSGS} introduced later in this section.

It is easy to prove concavity of the function $f$ (\ref{target function}).
\begin{proposition} \label{proposition concavity}
	Function $f$ defined in (\ref{target function}) is concave in $\Delta_s$.
\end{proposition}
\cite{Andrew1993} show that $f$ is differentiable at $w\in \Delta_s$ if and only if $f(w)$ is a simple eigenvalue of $\sqrt{\Qt} \Dt \sqrt{\Qt}$. It is also known that convex functions in Euclidean spaces are differentiable almost everywhere w.r.t. Lebesgue measure (see \cite{Borwein2010}, Section 2.5). \cite{Andrew1993} also provide exact formulas for computing derivatives of $f$ where they exist. Thus we are motivated to adapt subgradient method for convex functions in order to modify Step 3 in the above algorithm.

Let $<\cdot,\cdot>$ denote  scalar product in $\mathbb{R}^d$. Recall the definition of subgradient and subdifferential.
\begin{definition}
	Let $h: \mathbb{R}^d \to \mathbb{R}$ be a convex function. We say $v$ is a {\it subgradient} of $h$ at  point $x$ if for all  $y\in \mathbb{R}^d$,
	$$h(y)\geq h(x)+\<y-x,v\>.$$
	If $h$ is concave, we say that $v$ is a supergradient of $h$ at a point $x$, if $(-v)$ is a subgradient of the convex function $(-h)$ at $x$. The set of all $sub-(super-)gradients$ at the point $x$ is called {\it sub-(super-)differential} at $x$ and is denoted by $\partial h(x)$.
\end{definition}
In other word, $\partial h(x)$ parametrises a collection of all tangent hyperplanes at a point $x$.

Note that $f(w) =0$ on the boundary of $\Delta_{s}$. Therefore, the maximum of $f$ is  attained inside $\Delta_{s}$. One may apply the subgradient optimisation method in order to estimate $\po$. The method is described in Algorithm \ref{alg:subgradient method}.

\begin{algorithm} 
	\caption{Subgradient optimisation algorithm}\label{alg:subgradient method}
	Set an initial value of $w^0=(w^0_1,..,w^0_s) \in \Delta_{s}$. Define a sequence of non-negative numbers  $\(a_m\)_{m=1}^\infty$ such that $\sum_{m=1}^{\infty}a_m = \infty$ and $\lim_{m\to \infty} a_m = 0$. Set $i=0$.\\ 
	{\bf Beginning of the loop}
	\begin{enumerate}[label={\arabic*}.,ref={\arabic*}]
		\item Compute any $d^{i}\in \partial f(w^i)$. Normalise $d^i : = \frac{d^i}{|d^i_1|+..+|d^i_s|}$;
		\item $\wt_j := w_j^{i}+a_{i+1}d_j^{i}$, $j=1,..s$;
		\item  $w^{i+1} := \mathrm{Pr}_{\Delta_{s}}\(\wt\)$ , where $\mathrm{Pr}_{\Delta_{s}}$ is the projection operator on $\Delta_{s}$;
		\item $i:=i+1$.
	\end{enumerate}
	Go to {\bf Beginning of the loop} 
\end{algorithm}
\vspace{5mm}

It is known that Algorithm \ref{alg:subgradient method} produces a sequence $\{w^i\}$ such that $w^i \to w^\star$ as $i \to \infty$ (see Chapter 8 of \cite{Bertsekas2003}). Therefore, it is reasonable to combine the ARSGS \ref{alg:AGS general idea} with the subgradient algorithm. In order to do so, define a sequence of approximations of (\ref{target function}): 
$$\f(w)=\lm \(\Dnt    \Qnt \)=\lm \(\sqrt{\Qnt}  \Dnt    \sqrt{\Qnt} \).$$

\begin{algorithm}
	\caption{Adaptive Gibbs Sampler based on subgradient optimisation method (not implementable)}\label{alg:AGS subgradient}
	Generate a starting location $X_0\in \mathbb{R}^d$. Fix $\frac{1}{s+1}>\epsilon>0$. Set an initial value of $w^0=(w^0_1,..,w^0_s) \in \Delta^\epsilon_{s}$. Define a sequence of non-negative numbers  $\(a_m\)_{m=1}^\infty$ such that $\sum_{m=1}^{\infty}a_m = \infty$ and $\lim_{m\to \infty} a_m = 0$ . Set $i=0$.  Choose a sequence of positive integers $\(k_m\)_{m=0}^{\infty}$.\\
	{\bf Beginning of the loop}
	\begin{enumerate} [label={\arabic*}.,ref={\arabic*}]
		\item $n := n+k_i$. $p^i_j := \frac{w_j^i}{w^i_1+.. + w^i_s}$, $j=1,..,s$. Run RSGS($p^{i}$) for $k_{i}$ steps;
		\item Re-estimate $\Sn$ and $\Dnt$;
	\end{enumerate}		
	\begin{enumerate}[label=3.{\arabic*}.,ref=3.{\arabic*}]
		\item \label{alg:AGS subgradient:gradient}Compute $d^{i}\in \partial f_n(w^i)$.  Normalise $d^i := \frac{d^i}{|d^i_1|+..+|d^i_s|}$;
		\item \label{alg:AGS subgradient:move} $\wt_j := w_j^{i}+a_{i+1}d_j^{i}$, $j=1,..s$;
		\item \label{alg:AGS subgradient:compute projection} $w^{i+1} := \mathrm{Pr}_{\Delta_{s}^\epsilon}\(\wt\)$ , where $\mathrm{Pr}_{\Delta^\epsilon_{s}}$ is the projection operator on $\Delta^\epsilon_{s}$;
	\end{enumerate}	
	\begin{enumerate} [label*=\arabic*.]
		\setcounter{enumi}{3}
		\item $i:=i+1$.
	\end{enumerate}
	Go to {\bf Beginning of the loop} 
\end{algorithm}
\vspace{5mm}

Algorithm \ref{alg:AGS subgradient} resembles the aforementioned ideas. Here we consider iterations $w^{i}$ to be in $\Delta^\epsilon_{s}$ for $\epsilon>0$ because of  three reasons. Firstly, the RSGS with selection probabilities that are on the boundary of $\Delta_{s-1}$ is not ergodic. Secondly, this assumption is motivated by the results of \cite{Latuszynski2013}, where it is a minimum requirement to establish  convergence of  an Adaptive Gibbs Sampler. Finally, in the final Algorithm \ref{alg:ARSGS}, it is an essential assumption to be able to perform power iteration Step \ref{alg:ARSGS:power iteration}. Note, however, that $\epsilon >0$ may be chosen arbitrary small.

In order to construct an implementable and practical ARSGS algorithm, we still need to find a way to approximate the subgradient $ \partial f_n(w^i)$ in  Step \ref{alg:AGS subgradient:gradient}  and also find a cheap way of computing the projection $\mathrm{Pr}_{\Delta_{s}^\epsilon}\(\wt\)$ in Step \ref{alg:AGS subgradient:compute projection}.

An efficient algorithm to compute the projection on $\Delta_s^{\epsilon}$ is presented in \cite{Wang2013} and summarized in Algorithm \ref{alg:projection}. First, we increase all small coordinates to be $\epsilon$ in Step \ref{alg:projection:lift}. If the resulting point is outside $\Delta_s^{\epsilon}$, we need to project it on the hyperplane $\{w\in \mathbb{R}^s | 1 - \sum_{j=1}^{s}w = \epsilon,\ w_i\geq \epsilon\}$. In order to find the projection, we first rescale the coordinates in Step \ref{alg:projection:rescale}. Then we use the algorithm of \cite{Wang2013} to compute the projection on $\{w\in \mathbb{R}^s | \sum_{j=1}^{s}w = 1,\ w_i\geq 0\}$ in Steps \ref{alg:projection:sort} - \ref{alg:projection:lambda}. Finally, we rescale the resulting point in Step \ref{alg:projection:project} and thus obtain the desired projection.

\begin{algorithm}
	\caption{projection on $\Delta_{s}^\epsilon$} \label{alg:projection}
	The output of the algorithm is $w^{\rm proj}$ - projection of $w \in \mathbb{R}^s$ onto  $\Delta_{s}^\epsilon$.
	\begin{enumerate} [label={\arabic*}.,ref={\arabic*}]
		\item\label{alg:projection:lift} Define an auxiliary variable $w^{\rm aux} := w$. For $j=1,..,s$, if $w^{\rm aux} _j<\epsilon$, set $w^{\rm aux} _j: = \epsilon$;
		\item  If $1-\sum_{j=1}^{s}w^{\rm aux} _j> \epsilon$, then $w^{\rm proj} := w^{\rm aux}$ and go to  Step \ref{alg:projection:return}; else go to Step \ref{alg:projection:rescale};
		\item \label{alg:projection:rescale} For $j=1,..,s$, $w^{\rm temp} _j := \frac{1}{1-\epsilon (s+1)} (w^{\rm aux} _j-\epsilon)$;
		\item \label{alg:projection:sort} Sort vector $(w^{\rm temp} _1,..,w^{\rm temp} _{s})$ into $u:\ u_1\geq ...\geq u_{s}$;
		\item \label{alg:projection:rho}$\rho := \max \Bigg\{1\leq j\leq s:\ u_j +\frac{1}{j} \(1 -\sum_{k=1}^j u_k\)>0\Bigg\}$;
		\item \label{alg:projection:lambda} Define $\lambda =\frac{1}{\rho}\(1 -\sum_{k=1}^\rho u_k\)$;
		\item \label{alg:projection:project} For $j=1,..,s$,  $w^{\rm proj}_j := \epsilon + (1 - \epsilon (s+1)) \max\{w^{\rm temp} _j +\lambda, 0\}$; 
		\item \label{alg:projection:return} Return $w^{\rm proj}$.
	\end{enumerate}
\end{algorithm}
\vspace{5mm}

We are left to construct a procedure that approximates a supergradient $d^{i}\in \partial f_n(w^i)$ in Step \ref{alg:AGS subgradient:gradient} of Algorithm \ref{alg:AGS subgradient}. Since $\f(w)$ is the minimum eigenvalue of a self-adjoint matrix, $\f(w)$ may be obtained as
$$\f(w)=\min_{x:\|x\|=1}\<\sqrt{\Qnt} \Dnt \sqrt{\Qnt} x, x\>,$$
where $x\in \mathbb{R}^{d+1}$ and $\<\cdot,\cdot\>$ denotes scalar product in $\mathbb{R}^d$. Define
$$g^n_x(w) = \<\sqrt{\Qnt} \Dnt \sqrt{\Qnt} x, x\>.$$
Let $\nabla$  denote a gradient w.r.t. $w$. Then

{\footnotesize
	\begin{align}\label{gradient}
	\nabla g^n_x(w) = \(\<\sqrt{\Qnt} \frac{\partial \Dnt}{\partial w_1} \sqrt{\Qnt} x, x\>, .., \<\sqrt{\Qnt} \frac{\partial \Dnt}{\partial w_s}  \sqrt{\Qnt} x, x\>\).
	\end{align}
}

Here $\frac{\partial}{\partial w_i}$ stands for the element-wise derivative w.r.t. $w_i$, $i=1,..,s$.  Ioffe-Tikhomirov theorem (see, e.g., \cite{Zalinescu2002}) implies that the superdifferential of $\f$ at a point $w\in \Delta^\epsilon_{s}$ can be computed as

$$\partial \f(w)={\rm conv}\left\{\nabla g_x(w)\  \Bigg|\ x:\ \sqrt{\Qnt} \Dnt \sqrt{\Qnt} x = f_n(w) x,\ \|x\|=1 \right\},$$
where ${\rm conv} \{A\}$ denotes a convex hull of the set $A$.

Computing elements of the set  $\partial \f(w)$ is computationally expensive, since one has to calculate the minimum eigenvectors of $\sqrt{\Qnt} \Dnt \sqrt{\Qnt}$.  Therefore, we look for a cheap approximation of the points $\nabla g^n_x(w)$ in $\partial \f(w)$.

Let  $y = \sqrt{\Qnt}x$. Since we are interested in minimum eigenvectors $x$, such that
$$ \sqrt{\Qnt} \Dnt \sqrt{\Qnt} x = \f(w) x,$$
we can rewrite this equation as
\begin{align} \label{ascent direction}
\frac{1}{\f(w)} y = \(\Dnt\)^{-1} \Snt y.
\end{align}

That is, computing the minimum eigenvector of $\sqrt{\Qnt} \Dnt \sqrt{\Qnt}$ is equivalent to computing the maximum eigenvector of $ \(\Dnt\)^{-1} \Sn$. Given $y$ that solves (\ref{ascent direction}) and substituting  $x=\frac{1}{\|\sqrt{\Snt } y\|}\sqrt{\Snt}  y $ into (\ref{gradient}), we obtain

\begin{align} \label{gradient alternative in y}
\nabla g^n_x(w) = \frac{1}{\| \sqrt{\Snt } y\|^2} \Bigg( \Big< \frac{\partial \Dnt}{\partial w_1} y, y\Big> , ..,  \Big< \frac{\partial \Dnt}{\partial w_s} y, y \Big>\Bigg).
\end{align}

We can do further transformations.  Let 
\begin{align}\label{cholesky_decomp}
	\(\Dnt\)^{-1} = L_n (w) L_n^\mathrm{T} (w)
\end{align}

be the Cholesky decomposition of $\(\Dnt\)^{-1}$, where $L_n (w)$ is a lower triangular matrix. Define $z := L^{-1}_n(w) y$ and
\begin{align}\label{directional operator}
R_i (w) := \diag \(0,..,0,\frac{1}{w_i},..,\frac{1}{w_i},0,..,0,-\frac{1}{1-w_1 - .. - w_s}\),
\end{align}
where $\frac{1}{w_i}$ are placed exactly on the positions of the diagonal elements of $Q_{ii}$ in the partition  $Q = \(Q_{ij}\)_{i,j=1}^s$.
Then after simple manipulations, (\ref{ascent direction}) and (\ref{gradient alternative in y}) are equivalent respectively to
$$L_n^{\mathrm{T}}(w)\Snt L_n(w)z = \frac{1}{\f(w)} z $$
and 
\begin{align}\label{gradient alternative in z}
\nabla g^n_x(w) = \frac{1}{\< L_n^{\mathrm{T}}(w) \Snt L_n(w)z,  z\>}\Bigg( \< R_1(p) z, z\>, ..,\< R_{s-1}(p) z, z\>\Bigg),
\end{align}
where we used the block-diagonal structure of $L_n(w)$ and a representation
$$\frac{\partial \Dnt}{\partial w_i} = {\rm diag} \(0,.., 0, Q^{-1}_{ii}, 0,..,0,-1\).$$

Because of the normalisation in Step \ref{alg:AGS subgradient:gradient} of the Adaptive Gibbs Sampler \ref{alg:AGS subgradient},  (\ref{gradient alternative in y}) and (\ref{gradient alternative in z}) imply that a supergradient of $\f(w)$ is proportional to

\begin{align}\label{gradient alternative 1}
d_y (w) = \( \< (\Dnt)_1 y, y\>, .., \< (\Dnt)_{s-1} y, y\>\),
\end{align}
or, in terms of $z$, to
\begin{align}\label{gradient alternative 2}
d_z (w) = \( \< R_1 z, z\>, .., \< R_{s} z, z\>\),
\end{align}
where  $y$ and $z$ are the maximum eigenvectors of $ \(\Dnt\)^{-1} \Sn$ and\\ $L_n^{\mathrm{T}}(w)\Snt L_n(w)$, respectively. Here the lower triangular matrix $L_n (w)$ is defined by the Cholesky decomposition  (\ref{cholesky_decomp}).

Power iteration step may be performed in order to approximate $y$ and $z$. Let $z_0$ and $y_0$ be randomly generated unit vectors. Then at every iteration of the algorithm, we compute 
\begin{align}
&y_{i+1} = \Qnt D^{\rm ext}_n (w^i) y_i, \label{power iteration 1} \\
&z_{i+1} = L_n^{\mathrm{T}}(w^i)\Snt L_n(w^i) z_i.\label{power iteration 2}
\end{align}
and use the normalized vectors $y_{i+1}$ and $z_{i+1}$ instead of $y$ and $z$ when computing the directions (\ref{gradient alternative 1}) and (\ref{gradient alternative 2})

Given the intuition above, we present two versions of the ARSGS in the Algorithm \ref{alg:ARSGS}, where in round brackets we denote an alternative version of the algorithm.

One might notice the perturbation term $ b_{i+1} \xi_{i+1}$ in the Step \ref{alg:ARSGS:power iteration}. In fact, without the perturbation we may break the algorithm due to the fact that the power iteration step may fail to approach the maximum eigenvalue. It happens when $z_i$ (or $y_i$) "slips" into the eigenspace of a wrong eigenvalue and can't get out of it for the subsequent algorithm steps.

The simplest example one can think of is sampling from $N (0, I_2)$ using coordinatewise RSGS. Set $\Snt = I_3$. If one starts from $w^{0} = (\frac{1}{4}, \frac{1}{4}, \frac{1}{2})$ and steps $a_m$ (see Algorithm \ref{alg:ARSGS} for the meaning of $a_m$) are chosen to be tiny, $(0,0,1)$ is the maximum eigenvector of $ L_n^{\mathrm{T}}(w^i)\Snt L_n(w^i)= \diag (w^i_1,w^i_2, 1-w^i_1-w^i_1)$ for $i=1,..,N_0$, where $N_0$ depends on the sequence $a_m$. If $N_0$ is big enough (equivalently, $a_m$ is small enough), eventually $z_i = (0,0,1)$ for all $i$ due to the computational precision error and we will not get out of this eigenspace. Therefore, there is a possibility that eventually $w^{i}$ sticks to the boundary of $\Delta_{s}^\epsilon$. To surpass the issue we modify the power iteration Step \ref{alg:ARSGS:power iteration} by perturbing the values of $z_i$,
$$z_{i+1} = L_n^{\mathrm{T}}(w^i)\Snt L_n(w^i) z_i + b_i \xi_i,$$
where $b_i $ is a non-negative sequence convergent to 0, and $\xi_i$ is i.i.d. sequence of points uniformly distributed on the unit sphere.

\noindent {\bf Remark.} In Step \ref{alg:ARSGS:power iteration} of the ARSGS Algorithm \ref{alg:ARSGS} , $\frac{1}{\| L_n^{\mathrm{T}}(w^i)\Sn L_n(w^i) z_i \|}$ and $\frac{1}{\|\({D_n^{\rm ext}(w^{i})}\)^{-1} \Sn y_i\|}$ are approximations of $\underset{w\in \Delta_{s}^\epsilon}\max f(w)$, where $f$ is defined in (\ref{target function}). Therefore, taking into an account Proposition \ref{proposition extended pseudo-optimal points}, we can estimate $\PG(\po)$ by
\begin{align}\label{sp_gap:representation_1}
\PG(\po)\approx \((w^i_1+..+w^i_s)\ \| L_n^{\mathrm{T}}(w^i)\Snt L_n(w^i) z_i \|\)^{-1}
\end{align}
or 
\begin{align}\label{sp_gap:representation_2}
\PG(\po)\approx \((w^i_1+..+w^i_s)\  \|\({D_n^{\rm ext}(w^{i})}\)^{-1} \Snt y_i\|\)^{-1}.
\end{align}

\begin{algorithm}
	\caption{Adaptive Random Scan Gibbs  Sampler (final version)} \label{alg:ARSGS}
	Generate a starting location $X_0\in \mathbb{R}^d$. Fix $\frac{1}{s+1}>\epsilon>0$. Set an initial value of $w^0=(w^0_1,..,w^0_s) \in \Delta^\epsilon_{s}$,  generate a random unit vectors $z_0 \in \mathbb{R}^{d+1}$ (or $y_0 \in \mathbb{R}^{d+1}$). Define two sequences of non-negative numbers $\(b_m\)_{m=1}^\infty$ and $\(a_m\)_{m=1}^\infty$ such that $\sum_{m=1}^{\infty}a_m = \infty$, $a_m\to 0$ and $b_m \to 0$ as $m\to \infty$.  Set $i=0$.  Choose a sequence of positive integers $\(k_m\)_{m=0}^{\infty}$. \\
	{\bf Beginning of the loop}
	\begin{enumerate}[label={\arabic*}.,ref={\arabic*}]
		\item \label{alg:ARSGS:sample} $n := n+k_i$. $p^i_j := \frac{w_j^i}{w^i_1+.. + w^i_s}$, $j=1,..,s$. Run RSGS($p^{i}$) for $k_{i}$ steps;
		\item \label{alg:ARSGS:re-estimate} Re-estimate $\Sn$. Recompute $\Snt$, $D_n^{\rm ext}(w^{i})$, $ L_n(w^i)$;
	\end{enumerate}
	\begin{enumerate}[label={3.\arabic*}.,ref={3.\arabic*}]
		\item {\bf Compute approximate gradient direction $d^i$}:
		\begin{enumerate}[label={3.1.\arabic*}.,ref={3.1.\arabic*}]
			\item \label{alg:ARSGS:power iteration} Generate $\xi_{i+1}\sim N(0, I_{d+1})$. $\xi_{i+1}:=\frac{\xi_{i+1}}{\|\xi_{i+1}\|}$.\\
			Compute
			$$z_{i+1} := L_n^{\mathrm{T}}(w^i)\Snt L_n(w^i) z_i + b_{i+1} \xi_{i+1}.$$
			$$\Bigg(y_{i+1} :=\({D_n^{\rm ext}(w^{i})}\)^{-1} \Snt y_i + b_{i+1} \xi_{i+1}\Bigg)$$
			Normalise $z_{i+1}:= \frac{z_{i+1}}{\|z_{i+1}\|}.\ \Bigg(y_{i+1}:= \frac{y_{i+1}}{\|y_{i+1}\|}\Bigg).$
			
			\item  Compute  $d^{i} = d_{z_{i+1}} (w^{i})\ \mbox{ from } (\ref{gradient alternative 1})$ $\Bigg(d^{i} = d_{y_{i+1}} (w^{i}) \mbox{ from } (\ref{gradient alternative 2})\Bigg)$.
			Normalise $d^i : = \frac{d^i}{|d^i_1|+..+|d^i_s|}$;
		\end{enumerate}
		\item \label{alg:ARSGS:adapt weights}$\wt_j := w_j^{i}+a_{i+1}d_j^{i}$, $j=1,..s$;
		\item  Using Algorithm \ref{alg:projection} compute projection $w^{i+1}$ of $\wt$ onto $\Delta^\epsilon_{s}$;
	\end{enumerate}
	\begin{enumerate}[label={\arabic*}.,ref={\arabic*}]
		\setcounter{enumi}{3}
		\item \label{alg:ARSGS:iterate} $i:=i+1$.	
	\end{enumerate}
	Go to {\bf Beginning of the loop} 
\end{algorithm}

\section{Adapting Metropolis-within-Gibbs} \label{section AMWMwAG}
Sometimes one can not or does not want to sample from the full conditionals $Pr_i$ of the target distribution. In this case one may want to proceed with the Metropolis-within-Gibbs algorithm. For simplicity, we restrict ourselves to the coordinate-wise update Random Walk Metropolis-within-Gibbs (RWMwG) Algorithm \ref{alg:RWMwG}, though the idea presented below goes beyond this particular case.

One should not get confused with the parameter $q$ in Step \ref{alg:RWMwG:proposal} of the Algorithm \ref{alg:RWMwG}. If $q=1$, one recovers the RWMwG algorithm in its canonical form.

It is often not clear how to choose proposal variances $\beta_i$ to speed up the convergence.  We follow \cite{Gelman1996} suggestion that the average acceptance rate $\alpha$ should be around $0.44$ and adapt $\beta_i$ on the fly to keep up with this acceptance rate. Algorithm \ref{alg:ARWMwG} is the adaptive version of the RWMwG as suggested in \cite{Latuszynski2013}.

One could also adapt the selection probabilities $p_i$ but, as noted in \cite{Latuszynski2013}, there is no to-date guidance on the optimal choice of $p_i$. Heuristically, we would expect the Adaptive RWMwG to mimic the RSGS, so that we find it to be reasonable to adapt the selection probabilities $p_i$ in the same manner as for the RSGS. Therefore, we introduce Adaptive Random Walk Metropolis within Adaptive Gibbs  (ARWMwAG)  Sampler described in Algorithm \ref{alg:ARWMwAG}, where running the ARWMwG sampler in Step \ref{alg:ARWMwAG:run_ARWMwG} alternates with adaptation of the selection probabilities in Step \ref{alg:ARWMwAG:adapt_selecion_prob}.

\vspace{2mm}

\begin{algorithm}[H]
	\caption{Random Walk Metropolis-within-Gibbs (RWMwG)} \label{alg:RWMwG}
	Generate a starting location $X_0\in \mathbb{R}^d$. Let $(p_1, .., p_s)$ be a probability vector, $0< q\leq 1$, $\sigma^2>0$.  Fix variances $\beta_1,.. , \beta_s$ and choose starting location $\(X^0_1,.. , X^0_d\)\in \mathbb{R}^d$. $n:=0$.\\
	{\bf Beginning of the loop}
	\begin{enumerate}[label={\arabic*}.,ref={\arabic*}]
		\item \label{alg:RWMwG:select_coord} Sample $i\in \{1,.. ,s\}$ from probability distribution $(p_1, .., p_s)$;
		\item \label{alg:RWMwG:proposal} Draw $Y \sim  \left\{ \begin{array}{rcl}
		N(X_i, \beta^2_i) & \mbox{with probability}
		& q, \\
		N(X_i, \sigma^2)  & \mbox{with probability } &1-q;
		\end{array}\right. $
		\item \label{alg:RWMwG:acceptance} Compute  acceptance rate $\alpha = \min \Bigg\{1, \frac{\pi(Y | X^n_{-i})}{\pi(X^n_i | X^n_{-i})} \Bigg\};$
		\item \label{alg:RWMwG:update} With probability $\alpha$ accept the proposal  and set
		$$X^{n+1} = \(X^n_1, .. ,X^n_{i-1}, Y, X^n_{i+1},.., X^n_s\),$$
		otherwise, reject the proposal and set $X^{n+1} = X^n$;
		
		\item $n = n+1$.
		
	\end{enumerate}
	Go to {\bf Beginning of the loop} 
\end{algorithm}

\begin{algorithm}
	\caption{Adaptive Random Walk Metropolis-within-Gibbs (ARWMwG)} \label{alg:ARWMwG}
	Generate a starting location $X_0\in \mathbb{R}^d$. Let $(p_1, .., p_s)$ be a probability vector, $0 < q\leq 1$, $\sigma^2>0$.  Fix variances $\beta_1^0,.. , \beta_s^0$ and choose starting location $\(X^0_1,.. , X^0_d\)\in \mathbb{R}^d$. $n:=0$.\\
	{\bf Beginning of the loop}
	\begin{enumerate}[label={\arabic*}.,ref={\arabic*}]
		\item \label{alg:ARWMwG:sample}  Do Steps \ref{alg:RWMwG:select_coord} - \ref{alg:RWMwG:update} of RWMwG Algorithm \ref{alg:RWMwG} with proposal variances $(\beta_1^n,.. , \beta_s^n)$;
		\item \label{alg:ARWMwG:adapt} { $\beta_i^{n+1} = \beta_i^n \cdot \exp\(\frac{1}{n^{0.7}} (\alpha-0.44)\)$}, where $\alpha$ is the acceptance rate in Step \ref{alg:RWMwG:acceptance} of RWMwG Algorithm \ref{alg:RWMwG};
		\item $n = n+1$.
		
	\end{enumerate}
	Go to {\bf Beginning of the loop} 
\end{algorithm}

\begin{algorithm}[H]
	\caption{Adaptive Random Walk Metropolis within Adaptive Gibbs} \label{alg:ARWMwAG}
	Generate a starting location $X_0\in \mathbb{R}^d$. Fix variances $\beta_1^0,.. , \beta_s^0$, $0 < q\leq 1$, and $\sigma^2>0$. Choose also $\frac{1}{s+1}>\epsilon>0$. Set an initial value of $w^0=(w^0_1,..,w^0_s) \in \Delta^\epsilon_{s}$,  generate a random unit vector $z_0 \in \mathbb{R}^{d+1}$ (or $y_0 \in \mathbb{R}^{d+1}$). Define two sequences of non-negative numbers $\(b_m\)_{m=1}^\infty$ and $\(a_m\)_{m=1}^\infty$ such that $\sum_{m=1}^{\infty}a_m = \infty$, $a_m\to 0$ and $b_m \to 0$ as $m\to \infty$.  Set $i=0$.  Choose a sequence of positive integers $\(k_m\)_{m=0}^{\infty}$.\\
	{\bf Beginning of the loop}
	\begin{enumerate} [label={\arabic*}.,ref={\arabic*}]
		\item \label{alg:ARWMwAG:run_ARWMwG} $n := n+k_i$. $p^i_j := \frac{w_j^i}{w^i_1+.. + w^i_s}$, $j=1,..,s$. Iterate $k_{i}$ times Steps \ref{alg:ARWMwG:sample} and\\ \ref{alg:ARWMwG:adapt} of ARWMwG Algorithm \ref{alg:ARWMwG} with sampling weights $(p^{i}_1,..,p^{i},s)$ and proposal variances $(\beta_1^n,.. , \beta_s^n)$;
		\item \label{alg:ARWMwAG:adapt_selecion_prob} Do steps \ref{alg:ARSGS:re-estimate} - \ref{alg:ARSGS:iterate} of ARSGS Algorithm \ref{alg:ARSGS}. 
	\end{enumerate}
	Go to {\bf Beginning of the loop} 
\end{algorithm}

\section{Ergodicity of the Adaptive Gibbs Sampler}\label{section Ergodicity}

Here $\{P_\gamma\}_{\gamma\in\Gamma}$ is a collection of Markov kernels with a common stationary distribution $\pi$. For example, this can be a collection of RSGS kernels (\ref{kernel}) or the kernels of the Random Walk Metropolis Algorithm \ref{alg:RWMwG}.

The main result is presented in Theorem \ref{theorem ARGS convergence}, where ergodicity of the modified ARSGS Algorithm \ref{alg:ARSGS ergodic} is established under the {\it local simultaneous geometric drift} condition \ref{cond:local_simultaneous_drift}. We shall show in Theorem \ref{theorem local simultaneous drift} that the local simultaneous geometric drift is a natural condition for the ARSGS to have. More generally, if the condition  \ref{cond:local_simultaneous_drift} holds, we prove ergodicity for a class of modified AMCMC Algorithms \ref{alg:amcmc_modified}  in Theorem \ref{theorem amcmc_convergence}.

Ergodicity of the ARWMwG and ARWMwAG (Algorithms \ref{alg:ARWMwG} and \ref{alg:ARWMwAG}) is established under various conditions on the tails of the target distribution $\pi$ in Section 5 of \cite{Latuszynski2013}. In order to fulfil these conditions we, for example, could take arbitrary $0\leq q< 1$ and large enough $\sigma^2$ in the settings of the adaptive algorithms (see Theorems 5.6, 5.9 and Remark 5.8 in \cite{Latuszynski2013}).

For the ARSGS, we shall utilise Theorem 2 of \cite{Roberts2007}. The theorem guarantees ergodicity of an adaptive MCMC algorithm under the diminishing adaptation and containment conditions \ref{cond:diminishing} and \ref{cond:containment}.

\begin{enumerate}[label*=\bf  (C\arabic*)]
	\item\label{cond:diminishing} {\it Diminishing adaptation condition.} 
	$$\sup_{x\in \mathcal{X}}\|P_{\gamma_n}(x,\cdot) - P_{\gamma_{n+1}}(x,\cdot)\|_{TV} \to^{\rm P} 0\mbox{ as } n \to \infty,$$
	where $\|\cdot\|_{TV}$ is the total variation norm, $\gamma_n \in \Gamma$ - random sequence of parameters, and $\to^{\rm P}$ denotes the convergence in probability. Recall, for a signed measure $\mu$, its total variation norm $\|\mu\|_{TV} =  \sup_{A\in \mathcal{B} (\mathcal{X})}|\mu(A) |,$ where the supremum is taken over all measurable sets.
	\item\label{cond:containment} {\it Containment condition.} For $x\in \mathbb{R}^d$, $\gamma\in \Gamma$ and all $\epsilon>0$ define a function  
	$$M_\epsilon(x, \gamma):=\inf\Bigg\{N\geq 1\Bigg| \|P^N_{\gamma}(x,\cdot) - \pi(\cdot)\|_{TV}\leq \epsilon\Bigg\}.$$
	We say that an adaptive chain $\{X_n, \gamma_n\}$ satisfies the containment condition, if for all $\epsilon>0$, the sequence $\{M_\epsilon(X_n, \gamma_n)\}_{n=0}^\infty$ is bounded in probability, i.e.,  $\underset{N \to \infty}\lim \underset{n}\sup\ {\rm P}\Bigg(M_\epsilon(X_n,\gamma_n)>N\Bigg)  = 0$, where ${\rm P}$ is the probability measure induced by the chain.
\end{enumerate}

\noindent {\bf Theorem 2 of \cite{Roberts2007}.} {\it Let $\{X_n,\gamma_n\}$ be an 	  adaptive chain with $\{\gamma_n\}$ being the corresponding sequence of parameters. If $\{X_n, \gamma_n\}$ satisfies \ref{cond:diminishing} and \ref{cond:containment}, then the adaptive chain is ergodic, i.e., 	$$\|\mathcal{L} (X_n) - \pi\|_{TV} \to 0 \mbox{ as } n\to \infty,$$
where $\mathcal{L} (X_n)$ is the probability distribution law of $X_n$ and $\pi$ is the target distribution.}

One can easily see that \ref{cond:diminishing} holds for the ARSGS since $|p^{n+1} - p^{n}|\xrightarrow{P} 0$, due to the choice of decaying to zero adaptation rate $a_m$ in the Step \ref{alg:ARSGS:adapt weights} of Algorithm \ref{alg:ARSGS}.

Verifying the containment condition \ref{cond:containment} is less so trivial.  In Theorem 3 of \cite{Bai2009}, the containment is established if the {\it simultaneous geometric drift conditions} hold, i.e., if the following assumptions are fulfilled: 
\begin{enumerate}[label*=\bf  (A\arabic*)]
	\setcounter{enumi}{-1}
	\item \label{cond:simultaneous_minorisation} {\it Uniform small set.} There exist a uniform $(\nu_\gamma,m)-$small set $C$, i.e., there exists a measurable set $C\in \mathbb{R}^d$, an integer $m\geq 1$, a constant $\delta>0$ and a probability measure $\nu_\gamma$ probably depending on $\gamma\in \Gamma$, such that 
	\begin{align} \label{def:small_set}
	P^m_\gamma(x,\cdot)\geq \delta \nu_\gamma (\cdot).
	\end{align}
	\item\label{cond:simultaneous_drift} {\it Simultaneous geometric drift.} There exist numbers $b<\infty$, $0< \lambda < 1$, and a function $1\leq V< \infty$, such that $\underset{x\in C}{\sup} V(x) <\infty$ and for all $\gamma\in \Gamma,$
	$$P_\gamma V \leq \lambda V +b I_{C},$$
	where $P_\gamma V (x) = \underset{\mathbb{R}^d}\int V(y) P_\gamma (x, {\rm d} y)$ and the small set $C$ is defined in \ref{cond:simultaneous_minorisation}.
\end{enumerate}

Where the entire state space is small (i.e., $C = \mathbb{R}^d$ in \ref{cond:simultaneous_minorisation}) for some RSGS kernel $P_p$, $p\in \Delta_{s-1}^\epsilon$, ergodicity of the ARSGS is established in Section 4 of \cite{Latuszynski2013} (under additional $\pi-$irreducibility and aperiodicity assumptions).

In general, one could establish the simultaneous geometric drift condition \ref{cond:simultaneous_drift} and use Theorem 5.1 of \cite{Latuszynski2013} to derive the ergodicity. For the ARSGS, it might be hard to find a drift function $V$ that satisfies \ref{cond:simultaneous_drift}. Nevertheless, we  show that the {\it local simultaneous geometric drift condition} holds, provided that $P_p$ is geometrically ergodic for some $p\in \Delta_{s-1}^\epsilon$.

\begin{enumerate}[label*=\bf  (A\arabic*)]
	\setcounter{enumi}{1}
	\item \label{cond:geometric_drift} {\it Geometric ergodicity}. There exists $\gamma \in \Gamma$ such that $P_\gamma$ is geometrically ergodic. That is, $P_\gamma$ is $\pi-$irreducible, aperiodic (see Section 3.2 of \cite{Roberts2004} for definitions), and there exist drift coefficients  $(\lambda_\gamma, V_\gamma, b_\gamma, C_\gamma)$ such that 
	\begin{align}\label{def:geometric_drift}
	P_\gamma V_\gamma \leq \lambda_\gamma V_\gamma +b_\gamma I_{\{C_\gamma\}},
	\end{align}
	where $b_\gamma<\infty$, $0\leq \lambda_\gamma < 1$, $V_\gamma$ is a function such that $\pi-$almost surely $1\leq V_\gamma<\infty$ , and $ I_{\{C_\gamma\}}$ is an indicator function of a small set $C_\gamma$ (that is, for all $x\in C_\gamma$, (\ref{def:small_set}) holds).
\end{enumerate}

\begin{enumerate}[label*=\bf  (A\arabic*)]
	\setcounter{enumi}{2}
	\item \label{cond:local_simultaneous_drift} {\it Local simultaneous geometric drift.} For every $\gamma\in \Gamma$, there exists a measurable function $1\leq V_\gamma<\infty$, a small set $C_\gamma$ and an open neighbourhood $B_\gamma$ such that
	\begin{description}
		\item[\namedlabel{cond:local_simultaneous_drift:unif_set}{(a)}] $C_\gamma$ is a uniform small set for $\hat{\gamma} \in B_\gamma$, i.e., (\ref{def:small_set}) holds for all $\hat{\gamma} \in B_\gamma$ and $x\in C_\gamma$;
		\item[(b)] for all $\hat{\gamma} \in B_\gamma$,
		\begin{align}\label{local simultaneous drift}
		P_{\hat{\gamma}} V_\gamma \leq \tilde{\lambda}_\gamma V_\gamma +\tilde{b}_\gamma I_{\{C_\gamma\}}
		\end{align}
		for some $\tilde{b}_\gamma<\infty$ and $\tilde{\lambda}_\gamma<1$.
	\end{description}
	 
\end{enumerate}

\begin{theorem} \label{theorem local simultaneous drift}
	Assume \ref{cond:geometric_drift} for the RSGS kernels $P_p$, $p\in  \Delta_{s-1}^\epsilon =\Gamma $. Then $P_p$ is geometrically ergodic for each $p \in \Delta_{s-1}^\epsilon$ and satisfies the local simultaneous drift condition \ref{cond:local_simultaneous_drift}.
\end{theorem}

\noindent  {\bf Proof of Theorem \ref{theorem local simultaneous drift}.} 
Since for reversible $\pi-$irreducible chains, geometric ergodicity and existence of  $L_2-$spectral gap are equivalent (see Theorem 2 of \cite{Roberts2001}), the first statement follows from Theorem \ref{theorem upper bound}.

Let $(\lambda_p, V_p, b_p, C_p)$ be the drift conditions that satisfy (\ref{def:geometric_drift}). For every selection probability vecor $p = (p_1,..,p_s)$ let $m = m(p) = \underset{i\in\{1,..,s-1\}}{\min}p_i$.
Define norm $|p| =\underset{i\in\{1,..,s-1\}}{\max} \left| p_i\right|$ and take $\delta = \delta(p)>0$ such that $(1+(s-1)  \delta)\lambda_p \leq 1$. Set $\tilde{\lambda}_p = (1+(s-1) \delta)\lambda_p$. Then for every $\hat{p}$ such that $\left| \hat{p} - p \right| \leq m \delta$,

\begin{align*}
&P_{\hat{p}} V_p = \sum_{i=1}^s \hat{p}_i Pr_iV_p \leq \sum_{i=1}^s (p_i +(s-1) m \delta) Pr_iV_p\leq (1+(s-1) \delta)\sum_{i=1}^s p_i  Pr_iV_p = \\
& = (1 + (s - 1) \delta) P_p V_p \leq \tilde{\lambda}_p V_p + (1+(s-1) \delta)b_p I_{\{C_p\}},
\end{align*}
where we used representation (\ref{kernel}) for $P_p$ and the bound
$$|p_{s} - \hat{p}_{s-1}| \leq \sum_{i=1}^{s-1} |p_i - \hat{p}_i| \leq (s-1)m\delta. $$\\

We are left to show that the condition {\bf \ref{cond:local_simultaneous_drift:unif_set}} of \ref{cond:local_simultaneous_drift} is satisfied. Indeed, fix any probability vector  $p \in \Delta_{s-1}^\epsilon$. Since $C_p$ is a small set, $P^m_p (x,\cdot)\geq \delta_0 \nu(\cdot)$ for some $m\geq 1$, $\delta_0>0$, some probability measure $\nu$ and all $x\in C_p$. Then for all  $\hat{p}\in \Delta_{s-1}^\epsilon$,
$$P^m_{\hat{p}}(x,\cdot)\geq \(\frac{\epsilon}{\underset{i\in\{1,..,s\}}\max\ p_i}\)^m P_p^m(x,\cdot) \geq  \(\frac{\epsilon}{\underset{i\in\{1,..,s\}}\max\ p_i}\)^m \delta_0 \nu(\cdot),$$\\ 
whence the condition {\bf \ref{cond:local_simultaneous_drift:unif_set}} follows.\\
$\square$\\

In order to derive the ergodicity of the ARSGS, we will need the following crucial consequence of the assumption \ref{cond:local_simultaneous_drift}.

\begin{theorem}\label{theorem finite partition}
	Assume that $\Gamma$ is compact in some topology and that the collection of Markov kernels $P_\gamma$ satisfy \ref{cond:local_simultaneous_drift}. Then there exists a finite partition of  $\Gamma$ into $k$ sets $F_i$ such that 
	$$\cup_{i=1}^k F_i = \Gamma,$$
	and simultaneous geometric drift conditions \ref{cond:simultaneous_minorisation} and \ref{cond:simultaneous_drift} hold inside $ F_i$ with coefficients $(\lambda, V_i, b, C_i)$, where $0\leq \lambda<1$, $1\leq V_i < \infty$, $\pi-$a.s., $b<\infty$, and $C_i$ is the uniform small set for $\gamma\in F_i$.
\end{theorem}

\noindent {\bf Proof of Theorem \ref{theorem finite partition}.} Notice,

$$\Gamma\subset \cup_{\gamma\in \Gamma} B_\gamma,$$
where $B_\gamma$ is an open neighbourhood of $\gamma$ as in the assumption \ref{cond:local_simultaneous_drift}. 
Since every open coverage of a compact set $\Gamma$ has a finite subcoverage (see, e.g., Theorem 6.37 of \cite{Hewitt1965}), there exist a finite number of  $B_\gamma$ that cover $\Gamma$, say $B_{\gamma_1},..,B_{\gamma_k}$. Then one can take $\lambda:= \max\{\lambda_{\gamma_1},.., \lambda_{\gamma_k}\}$, $F_i: = B_{\gamma_i}$, $b = \max\{b_{\gamma_1},.., b_{\gamma_k}\}$, $C_i:= C_{\gamma_i} $.\\
$\square$\\

\begin{enumerate}[label*=\bf  (A\arabic*)]
	\setcounter{enumi}{3}
	\item \label{cond:drift_bounded} Assumption \ref{cond:local_simultaneous_drift} holds and for a chosen set $B\in \mathbb{R}^d$, and the corresponding drift functions $V_\gamma ,\gamma\in \Gamma$ are bounded on $B$, i.e.,\\  $\underset{x\in B} \sup V_\gamma(x) < \infty.$
\end{enumerate}

We are now ready to state the main ergodicity result.

\begin{theorem} \label{theorem amcmc_convergence}
	Fix a measurable set $B\subset \mathcal{X}$. Assume $\Gamma$ is compact in some topology and let $P_\gamma $, $\gamma\in \Gamma$ be a collection of $\pi-$irreducible, aperiodic Markov kernels with a common stationary distribution $\pi$. Consider an AMCMC Algorithm \ref{alg:amcmc_modified}, where the adaptations are allowed to take place only when the adaptive chain $\{X_n\}$ visits $B$. Let the conditions \ref{cond:diminishing}, \ref{cond:local_simultaneous_drift} and \ref{cond:drift_bounded} hold and assume that for a starting location $(X_0, \gamma_0)$ of the adaptive chain, $\E_{(X_0, \gamma_0)} V_{\gamma_0}(X_0)<\infty$, where $V_{\gamma_0}$ is the drift function for the initial kernel $P_{\gamma_0}$. Then the adaptive chain $\{X_n\}$ produced by the Algorithm \ref{alg:amcmc_modified} is ergodic.
\end{theorem}

\begin{algorithm}[H]
	\caption{Modified AMCMC} \label{alg:amcmc_modified}
	Set some initial values for $X_0\in \mathcal{X}$; $\gamma_0 \in \Gamma$; $\overline{\gamma}:=\gamma_0$; $k:=1$; $n:=0$. Fix any measurable set $B\subset \mathcal{X}$.\\
	{\bf Beginning of the loop}
	\begin{enumerate}[label={\arabic*}.,ref={\arabic*}]
		\item \label{alg:amcmc modified:sample} sample $X_{n + 1} \sim P_{\overline{\gamma}} (X_{n},\cdot)$;
		\item\label{alg:amcmc modified:precompute} given $\{X_0,  .., X_{n + 1}, \gamma_0, .., \gamma_{n}\}$ update $\gamma_{n+1}$ according to some adaptation rule;
		\item\label{alg:amcmc modified:update} If $X_{n+1}\in B$, $\overline{\gamma} := \gamma_{n+1}$.
	\end{enumerate}
	Go to {\bf Beginning of the loop} 
\end{algorithm}

\noindent {\bf Proof of Theorem \ref{theorem amcmc_convergence}}. Since we assume the diminishing adaptation condition \ref{cond:diminishing}, the proof follows once we establish the containment \ref{cond:containment}.

Theorem \ref{theorem finite partition} yields there exists a finite partition $\{F_i\}_{i=1}^k$ such that 
$$\cup_{i=1}^k F_i = \Gamma,$$
where simultaneous geometric drift conditions hold within every $F_i$ with some drift coefficients $(\lambda, V_i, b, C_i)$ (as in Theorem \ref{theorem finite partition}).

On $\Gamma$ define a function $r$ such that $r(p) = j$ if $\gamma\in F_j$. \ref{cond:drift_bounded} yields there exists $M<\infty$ such that  $V_i (x) < M$ for $x\in B$, $i =1, .., k$.

As in the proof of Theorem 3 of  \cite{Bai2009}, to verify the containment condition, it suffices to prove that 
$$\sup_n \E[V_{r(\gamma_n)}(X_n)] < \infty,$$
where hereafter $\E = \E_{(X_0, \gamma_0)}$ is the expectation with respect to the probability measure generated by the adaptive chain started from $(X_0, p^0)$.

Drift condition (\ref{local simultaneous drift}) implies

\begin{align*}
&\E\left[V_{r(\gamma_{n+1})} (X_{n+1}) | X_n = x, \gamma_n = \gamma\right] = \\
& = \E\left[V_{r(\gamma_{n+1})} (X_{n+1}) I_{\{X_{n+1} \notin B\}} | X_n = x, \gamma_n = \gamma\right] +\\
& +  \E\left[V_{r(\gamma_{n+1})} (X_{n+1}) I_{\{X_{n+1} \in B\}} | X_n = x, \gamma_n = \gamma\right] \leq \\
&\leq \E\left[V_{r(\gamma_{n+1})} (X_{n+1}) I_{\{X_{n+1} \notin B\}} | X_n = x, \gamma_n = \gamma\right] + M  =\\
& = \E\left[V_{r(\gamma_{n})} (X_{n+1}) I_{\{X_{n+1} \notin B\}} | X_n = x, \gamma_n = \gamma\right] + M \leq\\
&\leq \E\left[V_{r(\gamma)} (X_{n+1})  | X_n = x, \gamma_n = \gamma\right] + M =\\
& = P_{r(\gamma)} V_{r(\gamma)}(x) + M \leq \lambda V_{r(\gamma)} (x) +b +M,
\end{align*}
where in the first inequality we used the condition \ref{cond:drift_bounded} and in the last one we used the fact that $\gamma_{n+1} = \gamma_n$, if $ X_{n+1}\notin B$. Here $0<\lambda<1$ and $b<\infty$ are as in Theorem \ref{theorem finite partition}. Integrating out $\gamma_n$ and $X_n$ leads to

$$\E\left[V_{r(\gamma_{n+1})} (X_{n+1}) \right] \leq \lambda E\left[ V_{r(\gamma_n)} (X_n)\right] +b +M,$$
implying (see Lemma 2 of \cite{Roberts2007}),

$$\sup_n \E[V_{r(\gamma_n)}(X_n)] \leq \max\left\{\E V_{r(\gamma_0)}(X_0),\frac{b+M}{1-\lambda}\right\}<\infty.$$
 \\
$\square$\\

The reader can easily see that Theorem \ref{theorem amcmc_convergence} can be applied to a modified version of the ARSGS  Algorithm \ref{alg:ARSGS ergodic}, where the adaptations are allowed to happen only when the adaptive chain hits a set $B$ that satisfies \ref{cond:drift_bounded}. 

\begin{theorem}\label{theorem ARGS convergence}
	Fix a measurable set $B\in \mathbb{R}^d$. Consider an Adaptive Random Scan Gibbs Sampler (ARSGS) Algorithm \ref{alg:ARSGS ergodic} that produces a chain $\{ X_n\}$ for which the selection probabilities $p^{n+1}$  are allowed to be changed only if $X_{n+1} \in B$ (i.e., $p^{n+1} = p^{n}$ if $X_{n+1}\notin B$).\\
	Let the assumption \ref{cond:geometric_drift} hold. Then the assumption \ref{cond:local_simultaneous_drift} holds.\\
	Let also \ref{cond:drift_bounded} be satisfied for the set $B$ and assume that for the starting location $(X_0, p^0)$ of the adaptive chain, $\E_{(X_0, p^0)} V_{p^0}(X_0)<\infty$, where $V_{p^0}$ is the drift function for the initial kernel $P_{p^0}$. Then the adaptive chain $\{X_n\}$ produced by the ARSGS Algorithm \ref{alg:ARSGS ergodic} is ergodic. 
\end{theorem}

\noindent {\bf Proof of Theorem \ref{theorem ARGS convergence}.} Since $\Delta_{s-1}^\epsilon$ is closed and bounded, it is compact (see Heine-Borel Theorem in \cite{Hewitt1965}). Theorem \ref{theorem local simultaneous drift} implies that \ref{cond:local_simultaneous_drift} holds. The diminishing adaptation condition $\ref{cond:diminishing}$ holds since $\underset{i\in\{ 1,..,s\}}{\max}|p_i^{n+1} - p_i^n| \to 0 $ as $n \to \infty$, by the construction of the ARSGS Algorithm \ref{alg:ARSGS}. Therefore, we are in a position to apply Theorem \ref{theorem amcmc_convergence} to derive the desired ergodicity of the adaptive chain. \\
$\square$\\

\noindent {\bf Remarks} 
\begin{enumerate}[label*=\bf  \arabic*)]
	\item We do not have a proof that the ARSGS Algorithm \ref{alg:ARSGS} presented in Section \ref{section Adaptive Gibbs} is ergodic. However, the modified Algorithm \ref{alg:ARSGS ergodic} is ergodic under the assumptions of Theorem \ref{theorem ARGS convergence}. The only difference of the ergodic modification from the original version of the ARSGS is that we do not change the sampling weights $p_i$ if the chain is not in the set $B$. 
	\item The idea of introducing the set $B$ to an adaptive algorithm comes from the work of Craiu et al.\cite{Craiu2015}, where the authors study stability properties (e.g., recurrence) of adaptive chains where the adaptations are allowed to occur only in the set $B$.
	\item Assumption \ref{cond:drift_bounded} is satisfied for level sets $B = B(N) = \cap_{i=1}^k \{x: V_i(x) < N\}$, where $V_i$ are the drift functions as in the proof of the theorem. Theorem 14.2.5. of \cite{Meyn2009} implies that for large $N$, $B(N)$ covers most of the support of $\pi$, meaning that the adaptation will occur in most of the iterations of the Algorithm \ref{alg:ARSGS ergodic}.
	\item In practice often one can choose $B$ to be any bounded set in $\mathbb{R}^d$.
\end{enumerate}

\begin{algorithm}
	\caption{Adaptive Random Scan Gibbs  Sampler (ergodic modification)} \label{alg:ARSGS ergodic}
	Generate a starting location $X_0\in \mathbb{R}^d$. Fix a measurable set $B \subset \mathbb{R}^d$. Fix $\frac{1}{s+1}>\epsilon>0$. Set an initial value of $w^0=(w^0_1,..,w^0_s) \in \Delta^\epsilon_{s}$,  generate random unit vectors $z_0, y_0 \in \mathbb{R}^{d+1}$. Define two sequences of non-negative numbers $\(b_m\)_{m=1}^\infty$ and $\(a_m\)_{m=1}^\infty$ such that $\sum_{m=1}^{\infty}a_m = \infty$, $a_m\to 0$ and $b_m \to 0$ as $m\to \infty$.  Set $i=0$.  Choose a sequence of positive integers $\(k_m\)_{m=0}^{\infty}$.\\
	{\bf Beginning of the loop}
	\begin{enumerate}[label*=\arabic*.]
		\item $n:= n+k_i$. Run RSGS($p^{i}$) for $k_{i}$ steps;
		\item  Do Steps \ref{alg:ARSGS:re-estimate} - \ref{alg:ARSGS:iterate} of ARSGS Algorithm \ref{alg:ARSGS}.
		\item If current state of chain $X_n \in B$, then $p^i : = \frac{w^i}{w^i_1+.. + w^i_s}$;\\
		Otherwise, $p^{i} := p^{i-1}$.
	\end{enumerate}
	Go to {\bf Beginning of the loop} 
\end{algorithm}

\section{Simulations} \label{section simulations}

 It is known that for reversible Markov chains existence of the spectral gap is equivalent to geometric ergodicity (see Theorem 2 of \cite{Roberts2001}). Moreover, geometric ergodicity implies that the Central Limit Theorem holds (see, e.g.,  \cite{Bednorz2008}) . The following theorem  of Kipnis \& Varadhan \cite{Kipnis1986} states an important relation between the asymptotic variance in CLT and the spectral gap.
 
 \begin{theorem}\label{theorem asymptotic variance Kipnis-Varadhan}
 	Assume that $P_p$ is a RSGS kernel (\ref{kernel}). Then the following upper bound holds, connecting notions of the asymptotic variance with the spectral gap:
 	\begin{equation}\label{kipnis bound}
 	\sas(f) \leq \frac{2- \G(p)}{\G(p)} Var_\pi(f),
 	\end{equation}
 	where $Var_\pi$ denotes variance w.r.t. $\pi$ and $\G$ is the spectral gap of $P_p$. Moreover, if the spectrum of $P_p$ is discrete, then the equality in (\ref{kipnis bound}) is attained on a second largest eigenfunction of $P_p$.

 \end{theorem}
 
 Theorem \ref{theorem asymptotic variance Kipnis-Varadhan} states that by increasing the spectral gap one decreases the worst case asymptotic variance. Theorem \ref{theorem 2nd eigenfunction} states that the second largest eigenfunction of the RSGS kernel for the normal target distribution is a linear function. Of course, for arbitrary distribution  Theorem \ref{theorem 2nd eigenfunction} is false. Nevertheless, we believe that comparing the maximum asymptotic variance over linear functions for the adatptive and non-adaptive algorithms is a reasonable thing to do.  Define
 \begin{align}\label{linear_function}
 l_i = l_i(x) = \frac{x_i}{\sqrt{Var_\pi (x_i)}}
 \end{align}
 to be normalized  linear functions depending on one coordinate only.

 We compute the maximum asymptotic variance in CLT,  $\underset{i=1,..,d}{\max}\ \sas (l_i)$, for the adative and vanilla RSGS. We hope that in certain situations the ratio between the estimated pseudo-spectral gaps is close to the ratio of the maximum asymptotic variances over $l_i$ as follows from Theorem \ref{theorem asymptotic variance Kipnis-Varadhan}.  We study also how the pseudo-optimal weights affect the {\it autocorrelation function} (ACF) of $l_i$.

 Three different examples are studied where we implement coordinate-wise ARSGS, ARWMwAG and their non-adaptive versions. Two of the examples are in moderate dimension $50$: sampling from the posterior in a  Poisson Hierarchical Model (PHM) and sampling form the Truncated Multivariate Normal (TMVN) distribution. We also consider sampling from a posterior in a Markov Switching Model (MSM) in $200$-dimensional space.

 All the asymptotic variances are obtained using the batch-means estimator (see \cite{Galin2006} and \cite{Bednorz2007}). Below we outline settings for every problem.
 
 \subsection*{Poisson Hierarchical Model}

 Gibbs Sampler arises naturally for Hierarchical Models, where our goal is to sample from a posterior distribution. In the present model  data $Y_i$ comes from the Poisson distribution with intensity $\lambda_i$: 
 \begin{align}
 Y_i \sim {\rm Poisson} (\lambda_i), i=1,.., n,
 \end{align}
 where 
 \begin{align}
 \lambda_i  = \exp \(\sum_{j=1}^d x_{ij} \beta_j\),
 \end{align}
 with $\beta = (\beta_1,.. , \beta_d)$ being the parameter of interest. Stress that here $d$ is the dimensionality of the problem and  $n$ is the number of observations. We set $d=50$, $n=100$.
 
 Gibbs sampling through the adaptive rejection sampling presented by Gilks \& Wild in \cite{Gilks1992} is utilised for this problem. See \cite{Doss1994} for details and formulas for the full conditionals.
 
 We fix the true parameter $\beta_0 := (1,.. , 1)$ and take the prior distribution on $\beta$ to be normal with mean $(-1,.. , -1)$ and variance matrix $I_d$. We consider two different examples of the design matrix $X = {(x_{ij})_{i=1}^n}_{j=1}^d$.

 Design matrix $X^{(1)}$ is formed as follows. First, we set all the elements to be zero. Let $k = \frac{n}{d} = 2$. Then, we form two upper blocks of ones: $X^{(1)}_{ij} = 1$ for $i\in \{1,.., 2k\}$,  $j\in \{1, 2\}$,  and $X^{(1)}_{ij} = 1$ for $i \in \{2k+1,..,5k\}, \ j\in \{3,4,5\}$. Now there are at least  two blocks of correlated variables  in the posterior. For every other variable $\beta_j$, $j=5,.. , d$, set $X^{(1)}_{ij} = 1$ for $i\in \{ jk+1,.. , (j+1)k\}$.  In order to enforce dependency between all variables, we perturb every entry of $X^{(1)}$:  $X^{(1)}_{ij} = x^{(1)}_{ij} +0.1 \xi_{ij}$, where $\xi_{ij}$ are independent beta distributed variables with parameters $(0.1, 0.1)$.
 
 The second design matrix $X^{(2)}$ is formed as follows.  For $i =1,.. , n$ and $j = 1,.. , d$ set
 $$X^{(2)}_{ij} = 0.3 \( \delta_{ij}+\frac{\xi_i}{i}\),$$
 where $\delta_{ij}$ is the Kronecker symbol and $\xi_i$ are i.i.d.  beta distributed with parameters $(0.1, 0.1)$.
 
 \noindent {\bf Remark.} Correlation matrix of the posterior with the design matrix $X^{(1)}$ has blocks of highly correlated coordinates, whereas in case of the design matrix $X^{(2)}$, the correlation matrix seems to have only moderate non-diagonal entries.
 
 \subsection*{Truncated Multivariate Normal Distribution}
 
 Gibbs sampler is a natural algorithm to sample from the TMVN distribution as suggested by \cite{Geweke1991}. We consider linear truncation domain $b_1\leq Ax\leq b_2$ where $x\in \mathbb{R}^d$, $A$ is some $d\times d$ matrix. \cite{Geweke1991} suggested to transform the underlying normal distribution so that one needs to sample from $N(0,\Sigma_0)$ truncated to a rectangle $c_1\leq x\leq c_2$.
 
 We set $c_1 = (1,.. ,1) \in \mathbb{R}^{d}$, $c_2 = (3,.. , 3)\in \mathbb{R}^{d}$ and generate two different covariance matrices $\Sigma_0$. 
 
 $$\Sigma^{(1)}_0 ={\rm Corr}\(0.01 I_d + v_1 v_1^{\mathtt{T}}\),$$
 $$\Sigma^{(2)}_0 ={\rm Corr} \(0.01 I_d +   v_2 v_2^{\mathtt{T}}\),$$
 where $v_1 = (\xi_1, .., \xi_d)$, $v_2 = \(  \frac{\xi_1}{\log(2)}, \frac{\xi_2}{\log(3)} , .. ,  \frac{\xi_d}{\log(d + 1)}\)$, and $\xi_i$ are i.i.d. beta distributed with parameters $(0.1, 0.2)$. Here ${\rm Corr} (M)$ denotes a correlation matrix that corresponds to $M$.
 
 Note that the truncation domain does not contain the mode of the distribution, making it very different from the non-truncated normal distribution.c

 \subsection*{Markov Switching Model}

 Let $x_{1:i} = (x_1,..,x_i)$. We consider a version of stochastic volatility model where the underlying chain may be in either high or low volatility mode. Namely, the latent data $X_i$ forms an AR(1) process:
 $$X_{i} \sim N(X_{i-1}, \sigma^2_{r(i)}),$$
 where the chain can be in one of the two volatility regimes $r(i)\in \{0,1\}$. $r(i)$ itself forms a Markov chain with a transition matrix
 $$ \left( \begin{array}{cc}  1-a_1 & a_1  \\
 a_2&1-a_2
 \end{array} \right). $$
 Here $a_1$ and $a_2$ are called switching probabilities and assumed to be known. The observed data $Y_i$ is normally distributed: 
 $$Y_i \sim N(X_i, \beta^2)$$
 with known variance $\beta^2$. We consider data of $n=100$ observations and aim to sample from the posterior
 $$X_1,.., X_n, r(1),..,r(n) | Y_1,..,Y_n.$$
 
 Since $n=100$, the total number of parameters $d = 200$. We fix $a_1 = 0.001$ and $a_2 = 0.005$. \\
 
 The underlying hidden Markov chain $(X_i,r(i))$ is obtained as follows. We start chain $r(i)$ at it's stationary distribution and $X_0 \sim N(0, \sigma^2_{r(0)})$.  We then randomly generate chain $r(i)$ so that there are two switchings occur. Thus we obtain  $r(i) = 1$ for $i=57,.., 79$, and $r(i) = 0$, otherwise. 
 
 We consider data for 3 different combinations of $\sigma^2_0, \sigma^2_1$, and $\beta^2$:

 \begin{enumerate}[label=\bf (\alph*)]
 	\item \label{msm_a}  $\sigma^2_0=1$, $\sigma^2_1=10$,  $\beta^2=1$;
 	\item \label{msm_b}  $\sigma^2_0=1$, $\sigma^2_1=10$,  $\beta^2=3$;
 	\item \label{msm_c}  $\sigma^2_0=1$, $\sigma^2_1=5$,  $\beta^2=1$.
 \end{enumerate}
 
 Full conditionals may be obtained and are easy to sample from. We shall demonstrate performance of the coordinate-wise Gibbs Sampler for this problem.
 
 One might notice that the even and odd blocks of the coordinates can be updated simultaneously, meaning that coordinate-wise updates might be suboptimal. However, we are not motivated to find the best algorithms, but rather to demonstrate that the ARSGS can provide speed up even when the target distribution is discrete. Our intuition is such that the variables around the switching points will mix much slower, meaning that the ARSGS should update those coordinates more frequently.  
 
 \subsection{Adaptive Random Scan Gibbs Sampler} \label{section simulations ARSGS}

 We implement the ARSGS Algorithm \ref{alg:ARSGS} for all the examples. To do so, we need to specify a number of parameters in the algorithm. Since in the Euclidean space $\mathbb{R}^d$ distance from the origin to a simplex is $\frac{1}{\sqrt{n}}$, we find it reasonable to choose  $a_m = \frac{\log(50 \sqrt{d}+m)}{50 \sqrt{d}+m}$. In fact, one may choose arbitrary positive constant instead of $50$.  We do not know the right scaling for $b_m$ and thus set $b_m = a_m$.
 
 We set the lower bound $\epsilon:= \frac{1}{d^2}$. The choice of $\epsilon$ is  motivated by Theorem \ref{theorem upper bound}, where it is shown that the maximum improvement of the pseudo-spectral gap is $d$ (dimensionality of the space), which can only happen when one of the coordinates gets all of the probability mass. With the above choice of $\epsilon$, the maximum probability mass that coordinate can get is $1 - \frac{(d - 1)}{d^2}$, meaning that we might not be able to identify the optimal selection probabilities. On the other hand, the pseudo-spectral gap that corresponds to the selection probabilities obtained by the adaptive algorithm (with the specified value of $\epsilon = \frac{1}{d^2}$) will be close to the optimal value.
 
 We choose the sequence $k_m$ to be $k_m = 5000$ for PHM and TMVN examples, and $k_m = 8\times 10^5$ for MSM. We discuss an effective choice of $k_m$  in Section \ref{section computational cost}.

 We run the coordinate-wise ARSGS and the vanilla RSGS to obtain 5 million samples with 50 iterations thinning (i.e., we record every 50-th iteration of the chain)  in PHM and TMVN examples. Whereas, for the MSM example thinning is 8000 and number of samples is 10 million in cases \ref{msm_a}, \ref{msm_b}, and 30 million in the case \ref{msm_c}.

 \subsection*{Poisson Hierarchical Model}
 
 For this example, we show that in a long run the ARSGS not only outperforms the vanilla RSGS but performs similarly to the RSGS with the pseudo-optimal weights (that are estimated from the adaptive chain run).

 The data $Y_i$ is generated separately for the design matrices $X^{(1)}$ and $X^{(2)}$. We summarize results in Tables \ref{PHM Ex 1} and \ref{PHM Ex 2}, respectively. In a view of the von Mises theorem (see \cite{Vaart2000}), we expect the ARSGS to work well in both examples. One can observe reduction in the maximum asymptotic variance over the linear functions (\ref{linear_function}) by $9.27$ and $6.97$ times respectively.

 In the example with the design matrix $X^{(1)}$, the correlation between the 1st and 2nd coordinate is $-0.98$ and they are both nearly uncorrelated with the other coordinates. However, the corresponding optimal weights are such that $\po_1\approx \po_2 \approx 0.085$. The most of the probability mass is put on the coordinate $5$: $\po_5 \approx 0.29$. The maximum correlation of the coordinate $5$ with other directions is at most $0.57$ in absolute value. In fact, excluding coordinates 1, 2 and 5, all the off-diagonal correlations do not exceed $0.57$ in absolute value.

 For the second  example with the design matrix $X^{(2)}$, all the correlations are less than $0.21$. In some sense this example is consistent with the toy Example \ref{example2} from Section \ref{section example}. Here $\po_1\approx 0.188$, whereas all other optimal selection probabilities are in a range between $0.001$ and $0.01$.

\begin{figure}
 	\centering
 	\begin{subfigure}{\textwidth}
 		\centering
 		\includegraphics[scale=0.45]{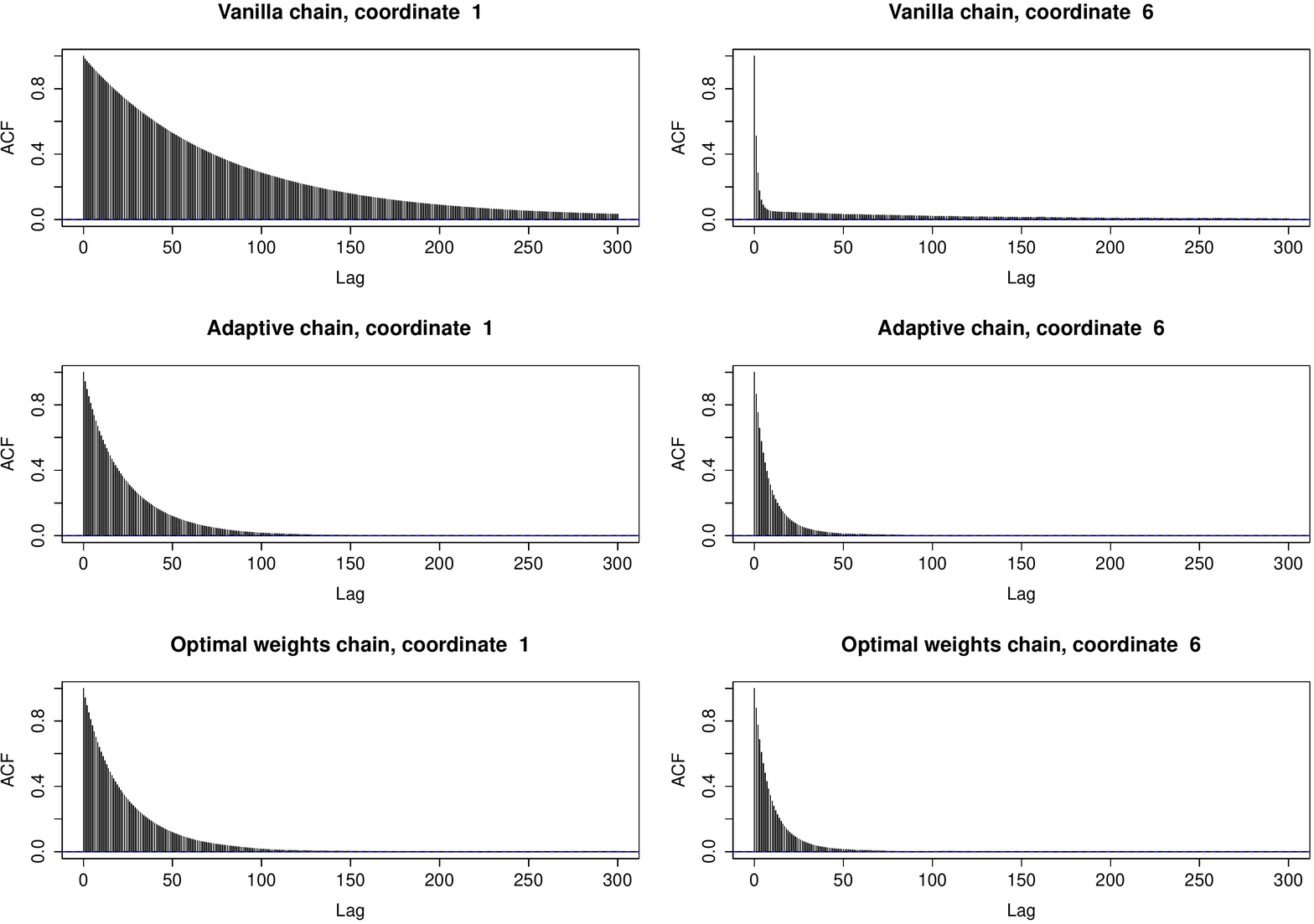}
 		\caption{Example 1}
 		\label{ACF PHM Ex 1}
 	\end{subfigure}%
 \\
 	\begin{subfigure}{\textwidth}
 		\centering
 		\includegraphics[scale=0.45]{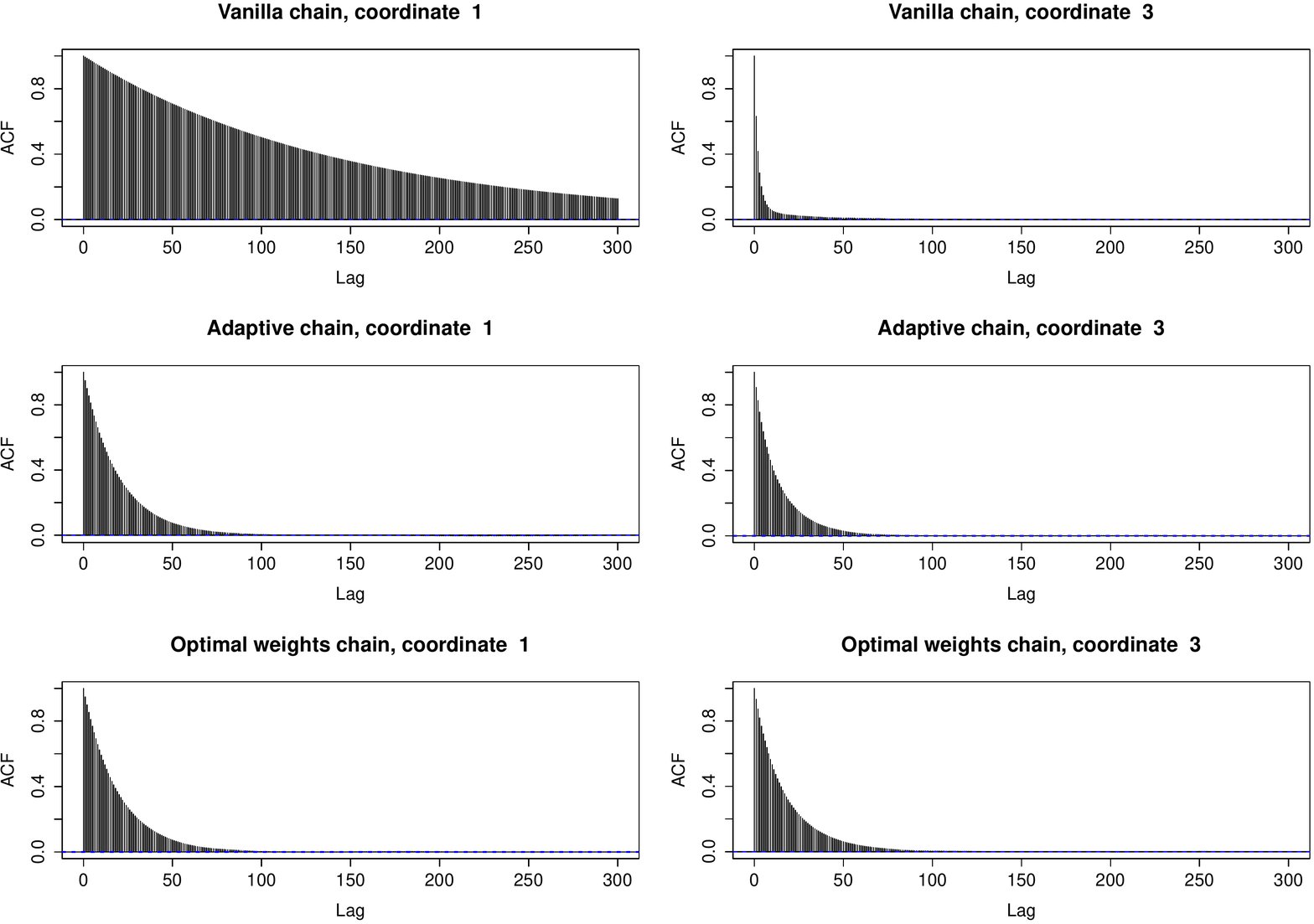}
 		\caption{Example 2} 
 		\label{ACF PHM Ex 2}
 	\end{subfigure}
 	\caption{d=50. PHM. ACF.}
 	\label{fig:test}
 \end{figure}

 Note that the RSGS with the optimal weights performs nearly the same  as the adaptive counterpart. For each design matrix ACF plots are produced for two coordinates with high and low optimal weights in Figures \ref{ACF PHM Ex 1} and 
 \ref{ACF PHM Ex 2}.

 Empirically, we observe that the adaptive algorithm tries to allocate the selection  probabilities in such a way that the effective number of independent samples (effective sample size) for every direction is the same. Hence, all the coordinates have about the same autocorrelation function (see Figure \ref{PHM Ex 1} and \ref{PHM Ex 2}). Note that the autocorrelation changes proportionally to the reduction in the  asymptotic variance.

 \pagebreak
 \begin{center}
 	\captionof{table}{PHM. Gibbs (d=50). Example 1}\label{PHM Ex 1}
 	\begin{tabularx}{0.9\textwidth}{*{7}{>{\centering\arraybackslash}X}}
 		\hline
 		& $1/\PG$ & $\underset{i}{\max}\ \av (l_i)$ \\ \hline
 		vanilla & 13435 &482 \\ \hline  
 		adaptive & 1355 & 52\\ \hline
 		optimal & - & 54\\ \hline
 		
 		$\frac{\mbox{vanilla}}{\mbox{adaptive}}$ & 9.9 &  9.27 \\ \hline
 	\end{tabularx}
 \end{center}
 \vspace{3mm}

 \begin{center}
 	\captionof{table}{PHM. Gibbs (d=50). Example 2}\label{PHM Ex 2}
 	\begin{tabularx}{0.9\textwidth}{*{7}{>{\centering\arraybackslash}X}}
 		\hline
 		& $1/\PG$ & $\underset{i}{\max}\ \av (l_i)$ \\ \hline
 		vanilla & 7340 &272 \\ \hline  
 		adaptive & 919 & 39\\ \hline
 		optimal & - & 40\\ \hline
 		
 		$\frac{\mbox{vanilla}}{\mbox{adaptive}}$ & 7.97 &  6.97 \\ \hline
 	\end{tabularx}
 \end{center}
 
 \nopagebreak

 {\bf Remark.} We use the Adaptive Rejection Sampling algorithm (see \cite{Gilks1992}) in order to sample from the full conditionals. Since the normalising constant is not known in this case, we could not establish geometric ergodicity of the RSGS in this case. On the other hand, results of Latuszynski et al. \cite{Latuszynski2013}  ensure that the RWMwG is geometrically ergodic. Since typically the RSGS converges faster than the corresponding RWMwG, we suggest that the RSGS is also geometrically ergodic for the Poisson Hierarchical Model. This means that heuristically the modified ARGS Algorithm \ref{alg:ARSGS ergodic} is ergodic in the current settings.

 \subsection*{Truncated Multivariate Normal Distribution}

\begin{figure}
 	\centerline{\includegraphics[scale=0.41]{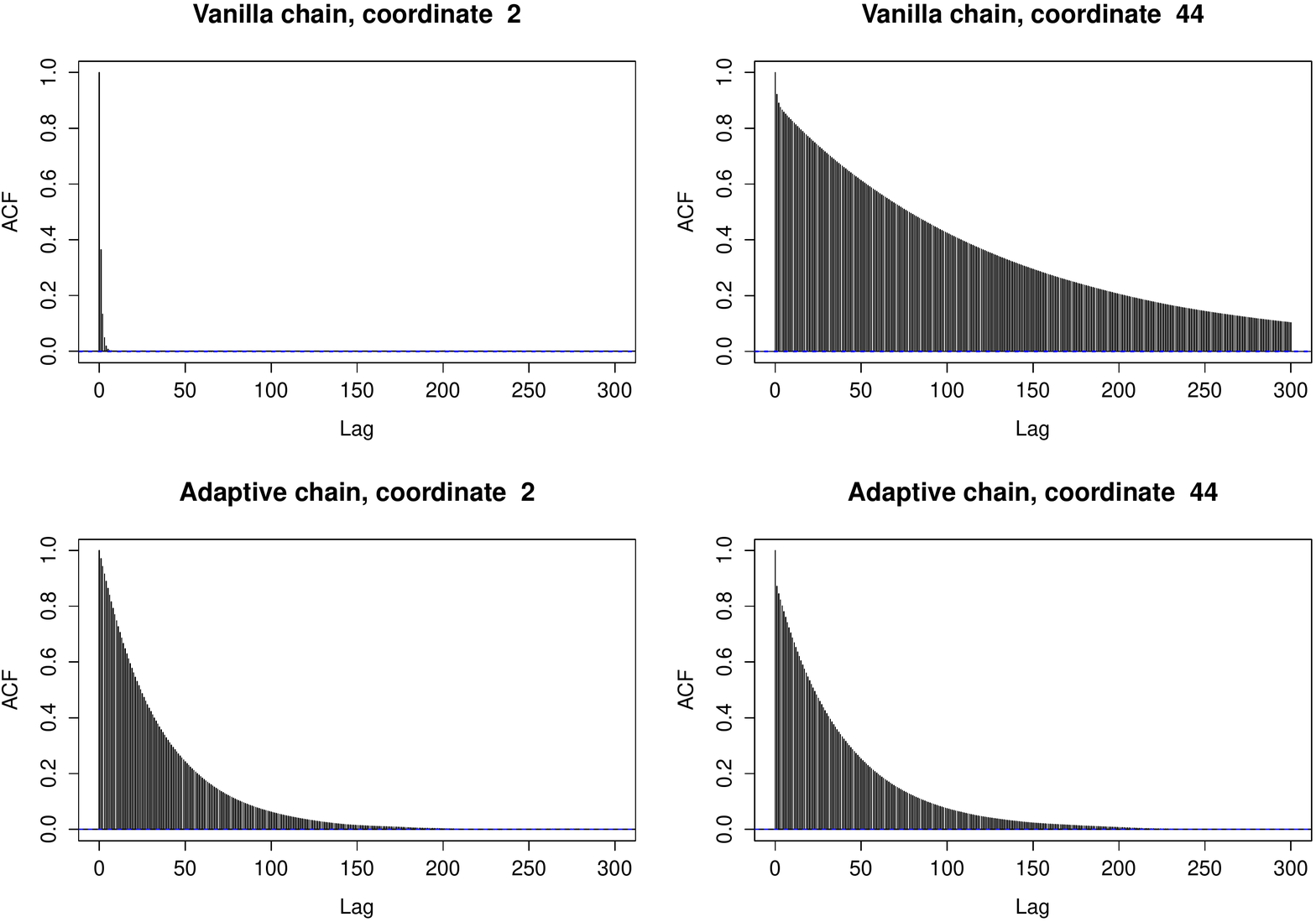}}
 	\caption{$d=50$. TMVN. ACF. Example 1} \label{ACF TMVN Ex 1}
 \end{figure}

 For the first correlation matrix  $\Sigma_0^{(1)}$ the reduction in the maximum asymptotic variance over the linear functions(\ref{linear_function}) is $3.44$, which is surprisingly very close to the ratio of the  pseudo-spectral gaps (Table \ref{TMVN Ex 1}). The autocorrelation plot of 2nd and 47th coordinates is in Figure \ref{ACF TMVN Ex 1}. The same effect of keeping the same autocorrelations for all the coordinates is observed.

 \pagebreak
 \begin{center}
 	\captionof{table}{TMVN ($d=50$). Example 1}\label{TMVN Ex 1}
 	\begin{tabularx}{0.9\textwidth}{*{7}{>{\centering\arraybackslash}X}}
 		\hline
 		& $1/\PG$ & $\underset{i}{\max}\ \av (l_i)$ \\ \hline
 		vanilla & 6449 &239 \\ \hline  
 		adaptive & 1857 & 72\\ \hline 
 		
 		$\frac{\mbox{vanilla}}{\mbox{adaptive}}$ & 3.47 &  3.32 \\ \hline
 	\end{tabularx}
 \end{center}
 \vspace{5mm}

 For the second correlation matrix $\Sigma_0^{(2)}$, the improvement of the asymptotic variance is only half of the improvement of the spectral gap, as seen from Table \ref{TMVN Ex 2}. However, we still observe that the ARSGS assigns more weight to the coordinates that mix slower.  
 
 \pagebreak
 \begin{center}
 	\captionof{table}{TMVN (d=50). Example 2}\label{TMVN Ex 2}
 	\begin{tabularx}{0.9\textwidth}{*{7}{>{\centering\arraybackslash}X}}
 		\hline
 		& $1/\PG$ & $\underset{i}{\max}\ \av (l_i)$ \\ \hline
 		vanilla & 467 &12.6  \\ \hline  
 		adaptive & 161 & 8.3 \\ \hline
 		
 		$\frac{\mbox{vanilla}}{\mbox{adaptive}}$ & 2.9 &  1.5 \\ \hline
 	\end{tabularx}
 \end{center}
 \vspace{5mm}
 
 {\bf Remark.}  In this example, the RSGS kernels (\ref{kernel}) satisfy the uniform minorisation condition \ref{cond:simultaneous_minorisation}. The corresponding small set is the whole domain, because it is compact. The diminishing adaptation condition \ref{cond:diminishing} holds by construction. Therefore, the ARSGS algorithm is ergodic by virtue of Theorem 4.2 \cite{Latuszynski2013}.

 \subsection*{Markov Switching Model}
 
 Note that half of the coordinates in the target distribution are discrete. The naive estimator (\ref{naive estimator}) of the covariance structure is often singular (i.e., non-invertible) in these settings. Therefore, to implement the ARSGS, in Step \ref{alg:ARSGS:re-estimate} of the Algorithm \ref{alg:ARSGS}, we use a perturbed naive estimator $\Sn+\frac{1}{d^3} I_d$, where $d=200$ is dimensionality of the target distribution. We did not observe any significant impact of the added perturbation on the estimated optimal selection probabilities.  
 
 We expect that if the dependency structure is described by the correlations (i.e., zero correlation implies weak in some sense dependency), then the ARSGS shall outperform the vanilla RSGS. We observe that this happens in cases  \ref{msm_a} and \ref{msm_b}. More precisely, the ARSGS tends to put more weight on coordinates that have larger asymptotic variance and results are found in Table \ref{MSM Gap}, where the improvement of the pseudo spectral gap for each case is presented, and  in Table \ref{MSM asymptotic}, where the corresponding improvement in asymptotic variance over the linear functions (\ref{linear_function}) is presented.  Notice, in all cases the maximum asymptotic variance $\underset{i}{\max}\ \av (l_i)$ is attained for the coordinate $i$ that corresponds to some discrete direction $r(i)$.
 \pagebreak
 \begin{center}
 	\captionof{table}{MSM (d=200). $1/\PG$}\label{MSM Gap}
 	\begin{tabularx}{0.9\textwidth}{*{7}{>{\centering\arraybackslash}X}}
 		\hline
 		&  {\bf (a)} & {\bf (b)} & {\bf (c)} \\ \hline
 		vanilla & 18875 & 71106 & 117127\\ \hline  
 		adaptive  & 3450 & 9924 &  19301\\ \hline
 		
 		$\frac{\mbox{vanilla}}{\mbox{adaptive}}$ & 5.47 &  7.17  & 6.07  \\ \hline
 	\end{tabularx}
 \end{center}
 
 \begin{center}
 	\captionof{table}{MSM (d=200). $\underset{i}{\max}\ \av (l_i)$}\label{MSM asymptotic}
 	\begin{tabularx}{0.9\textwidth}{*{7}{>{\centering\arraybackslash}X}}
 		\hline
 		&  {\bf (a)} & {\bf (b)} & {\bf (c)} \\ \hline
 		vanilla & 21.7 & 45.7 & 197   \\ \hline  
 		adaptive & 6  & 12.6 &  204 \\ \hline
 		
 		$\frac{\mbox{vanilla}}{\mbox{adaptive}}$ & 3.6 & 3.63  & 0.97 \\ \hline
 	\end{tabularx}
 \end{center}
 \vspace{5mm}
 
 The case \ref{msm_c} is special in a sense that the correlation structure does not reveal the dependency structure. Here the maximum selection probability is assigned to the coordinate that corresponds to the variable $r(99)$. The asymptotic variance for the corresponding linear function $l_{i}$  drops roughly by $4.5$ times from $40.56$ to $9.06$.  However, the  maximum asymptotic variance over the linear functions (\ref{linear_function}) is attained on the coordinate $i$ that corresponds to $r(64)$.  The estimated  optimal weight $\po_i$ corresponding to $r(64)$ is only slightly larger than the uniform weight $1/200$.

 To justify convergence of the ARSGS and the results in Tables \ref{MSM Gap} and \ref{MSM asymptotic}, we provide a proof of the geometric ergodicity.
 
 \begin{proposition}\label{proposition ergodicity MSM}
 	In cases \ref{msm_a}, \ref{msm_b} and \ref{msm_c}, the RSGS for the Markov Switching Model described above is geometrically ergodic, i.e., satisfies \ref{cond:geometric_drift}.
 \end{proposition}

 \subsection{Adaptive Random Walk Metropolis within Adaptive Gibbs} \label{section simulations ARWMwAG}
 
 As before, we sample from the same PHM in $\mathbb{R}^{50}$. We consider the same algorithm settings for the ARWMwAG Algorithm \ref{alg:ARWMwAG} as for the ARSGS Algorithm \ref{alg:ARSGS}. We compare performance of the Random Walk Metropolis-within-Gibbs algorithm with it's adaptive versions ARWMwG and ARWMwAG.
  
 For the RWMwG the proposal variances $\beta_i$ are chosen to be ones. For demonstration purposes, the parameter $q$ of the adaptive versions of the algorithm, is  chosen to be $q:=1$.

 \subsection*{Poisson Hierarchical Model}
 
\begin{figure}
 	\centering
 	\begin{subfigure}{\textwidth}
 		\centering
 		\includegraphics[scale=0.45]{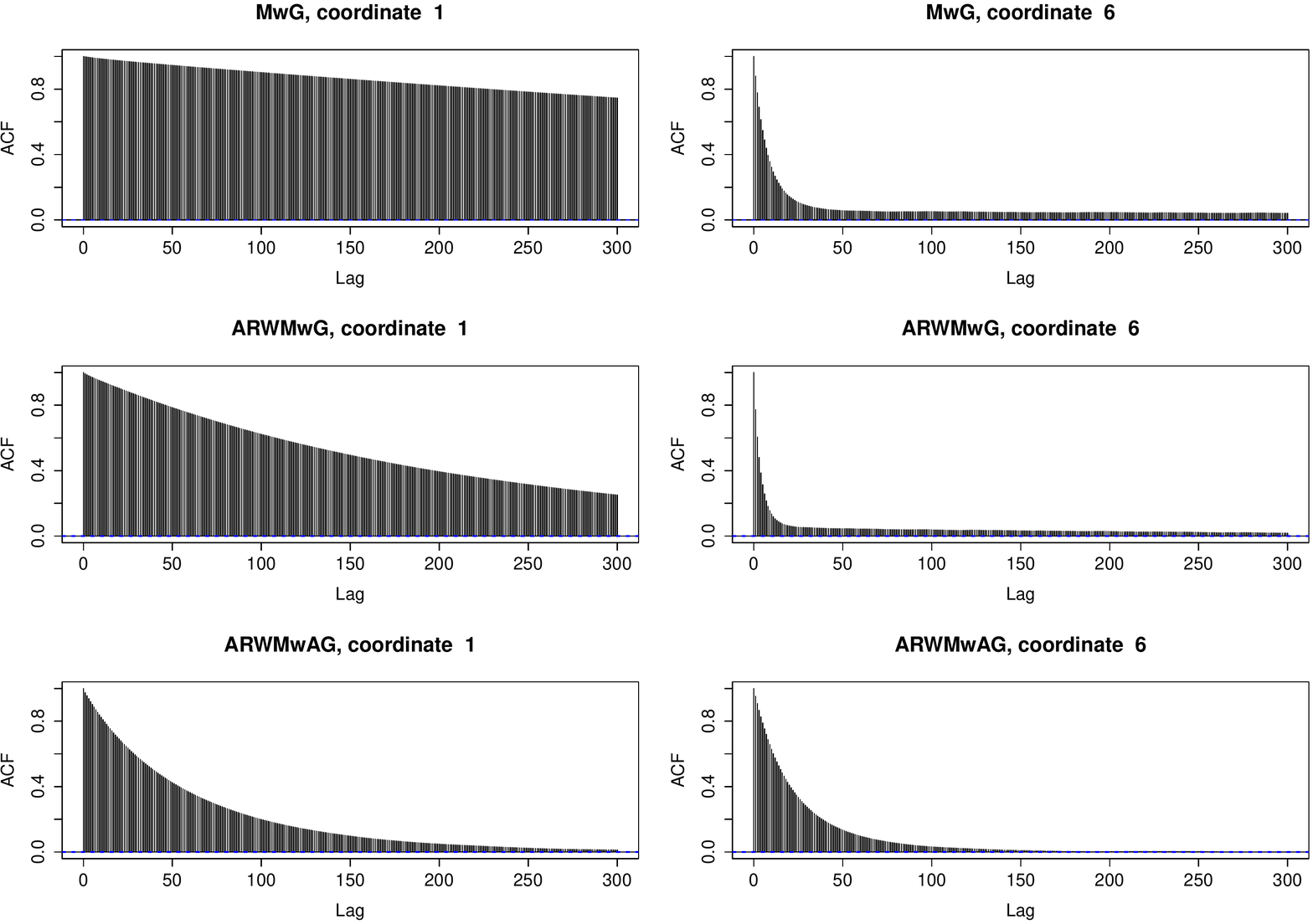}
 		\caption{Example 1}\label{ACF MwG PHM Ex 1}
 	\end{subfigure}%
\\
 	\begin{subfigure}{\textwidth}
 		\centering
 		\includegraphics[scale=0.45]{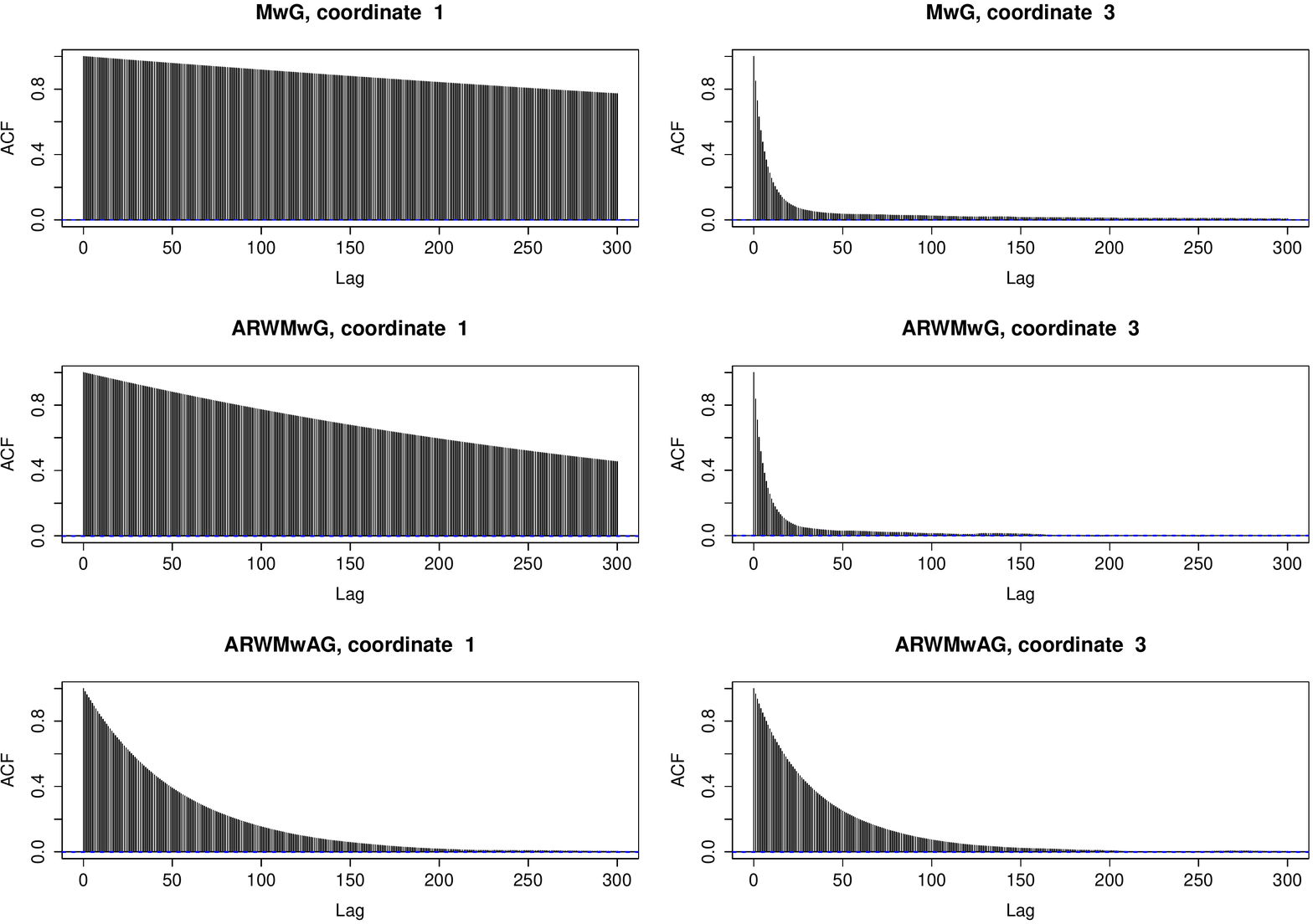} 
 		\caption{Example 2} \label{ACF MwG PHM Ex 2}
 	\end{subfigure}
 	\caption{d=50. PHM. ACF.}
 	\label{fig:ACF_MwG}
 \end{figure}

 Tables \ref{PHM MwG Ex 1} and \ref{PHM MwG Ex 2} provide the analysis of the asymptotic variances. For the first design matrix $X^{(1)}$, we observe a $7$ times improvement of the ARWMwAG over the ARWMwG algorithm, and the total improvement of almost $15$ times over the non-adaptive RWMwG. For the second design matrix $X^{(2)}$, the corresponding  improvement is $6.1$ and $12.3$ times respectively.  On Figure \ref{fig:ACF_MwG} we present the improvements to the ACF. 

 \begin{center}
 	\captionof{table}{PHM. MwG (d=50). Example 1}\label{PHM MwG Ex 1} 
 	\begin{tabularx}{0.9\textwidth}{*{7}{>{\centering\arraybackslash}X}}
 		\hline
 		& $1/\PG$ & $\underset{i}{\max}\ \av (l_i)$\\ \hline
 		RWMwG (non-adaptive)&  -- & 1993 \\ \hline
 		ARWMwG (partially adaptive) &13244 &  971 \\ \hline
 		ARWMwAG \linebreak (fully adaptive)&1376  & 138\\ \hline
 		$\frac{\mbox{partially adaptive}}{\mbox{fully adaptive}}$& 9.63  & 7 \\ \hline
 		$\frac{\mbox{non-adaptive}}{\mbox{fully adaptive}}$& -- & 14.45  \\ \hline
 	\end{tabularx}
 \end{center}
 \vspace{5mm}
 \pagebreak
 \begin{center}
 	\captionof{table}{PHM. MwG (d=50). Example 2}\label{PHM MwG Ex 2}
 	\begin{tabularx}{0.9\textwidth}{*{7}{>{\centering\arraybackslash}X}}
 		\hline
 		& $1/\PG$ & $\underset{i}{\max}\ \av (l_i)$\\ \hline
 		RWMwG (non-adaptive)& -- & 1276 \\ \hline
 		ARWMwG (partially adaptive) & 7461 &639 \\ \hline
 		ARWMwAG \linebreak (fully adaptive)& 970 & 104\\ \hline
 		$\frac{\mbox{fully adaptive}}{\mbox{partially adaptive}}$&  7.69 & 6.14 \\ \hline
 		$\frac{\mbox{fully adaptive}}{\mbox{non-adaptive}}$& -- & 12.27  \\ \hline
 	\end{tabularx}
 \end{center}
 \vspace{5mm}

 {\bf Remark.} The target distribution satisfies the Assumption 5.4 of \cite{Latuszynski2013}. If one chooses the proposal in Step \ref{alg:RWMwG:proposal} of RWMwG Algorithm \ref{alg:RWMwG} to be a mixture of normals, i.e., $0<q<1$, or restricts the proposal variances $\beta_i$ to be in some interval $[c_1,c_2]$, $\infty>c_2>c_1>0$, then the ARWMwG and ARWMwAG is ergodic by the virtue of Theorem 5.5 of \cite{Latuszynski2013}.
 
 \subsection{Computational cost of the adaptation} \label{section computational cost}
 
 By doing the adaptations of an MCMC algorithm we increase the total running time of the algorithm. Complexity of the projection Algorithm \ref{alg:projection} is bounded by the complexity of a sorting algorithm used in Step \ref{alg:projection:sort}, which is usually of order $\mathcal{O}(d \log(d))$. Thus one can easily see that the total adaptation cost of Steps \ref{alg:ARSGS:re-estimate} - \ref{alg:ARSGS:iterate}  of the ARSGS Algorithm \ref{alg:ARSGS} is bounded by the complexity of the Step \ref{alg:ARSGS:re-estimate}, which requires finding the diagonal blocks of the inverted covariance matrix. Usual Gauss matrix inversion is of order $\mathcal{O}(d^3)$. In high dimensional settings it is an expensive procedure. However, one can choose the sequence $k_m$ in the setting of the ARSGS Algorithm \ref{alg:ARSGS} in order to make the adaptation cost negligible comparing to the sampling Step \ref{alg:ARSGS:sample}.

 Turn back to the Poisson Hierarchical Model example with the design matrix $X^{(1)}$. The sequnce $k_m$ was chosen to be $k_m = 5000$. In column 2 of Table \ref{PHM Ex 1 cost} we put the average real time in seconds spent  on sampling Step \ref{alg:ARSGS:sample}  of the  ARSGS and ARWMwAG algorithms. The average time spent for one adaptation (i.e., to perform Steps \ref{alg:ARSGS:re-estimate} - \ref{alg:ARSGS:iterate}  of the ARSGS Algorithm \ref{alg:ARSGS} is in column 3. The maximum asymptotic variance over the linear functions (\ref{linear_function}) is in column 1.
 
 \nopagebreak
 \begin{center}
 	\captionof{table}{PHM. Example 1 (d=50)}\label{PHM Ex 1 cost}
 	\begin{tabularx}{0.9\textwidth}{*{7}{>{\centering\arraybackslash}X}}
 		\hline
 		& $\underset{i}{\max}\ \av (l_i)$ & Cost per 5000 iterations & Cost of adaptation \\ \hline
 		ARSGS  & 52 & 0.37  &  0.0025\\ \hline
 		ARWMwAG &138 & 0.028 &   0.0025 \\ \hline
 	\end{tabularx}
 \end{center}
 \vspace{2 mm}

 Gibbs Sampling for the PHM requires the use of the adaptive rejection sampling (see \cite{Doss1994, Gilks1992}), which significantly increases the time needed to obtain a sample. Therefore, even though the ARSGS has $2.65$ times lower asymptotic variance than the ARWMwAG algorithm, it samples more than $10$ times slower.
 
 By adjusting the sequence $k_m$ one can tune the ratio of the adaptation time over the sampling time. In fact, the sampling and adaptations can be performed  independently in a sense that they may be computed on different CPUs as demonstrated in the Algorithm \ref{alg:parallel}.

 \begin{algorithm}
 	\caption{Parallel versions of ARSGS and ARWMwAG} \label{alg:parallel}
 	Set all initial parameters for the ARSGS $\Big($ARWMwAG $\Big)$.\\
 	{\bf Do on different CPUs:}
 	\begin{itemize}
 		\item $n = n+k_i$. $p^i = \frac{w^i}{w^i_1+.. + w^i_s}$ Run  RSGS($p^{i}$) $\Big($or ARWMwG($p^{i}$)$\Big)$ for $k_{i}$ steps.
 		\item Do steps the steps  2 - 4 of ARSGS based on available chain output.
 	\end{itemize}
 	{\bf Wait till both CPUs finish their jobs. Then iterate the procedure.} 
 \end{algorithm}

\section{Discussion} \label{section discussion}

We have devised the Adaptive Random Scan Gibbs and Adaptive Random Walk Metropolis within Adaptive Gibbs algorithms, where adaptations are guided by optimising the $L_2-$spectral gap for the Gaussian target analogue called pseudo-spectral gap. The performance of the adaptive algorithms has been studied in Section \ref{section simulations}. We have seen that it might hard to decide in advance whether the adaptive algorithm would outperform the non-adaptive counterpart. On the other hand, as suggested in Section  \ref{section computational cost}, the computational time added by the adaptation can be made negligible comparing to the total run time of the algorithm. Therefore, we believe that it is reasonable to utilise the adaptive algorithms given that substantial computational gain may be achieved. However, one needs a natural notion of the covariance structure for the target distribution in order to implement the adaptive algorithms.

We have analysed ergodicity property of the adaptive algorithms in Section \ref{section Ergodicity}. We have developed a concept of the local simultaneous drift condition \ref{cond:local_simultaneous_drift}. We have shown in Theorem \ref{theorem local simultaneous drift} that the condition is natural for the ARSGS. Under this condition, in Theorem \ref{theorem amcmc_convergence} we have established ergodicity of modified AMCMC Algorithms \ref{alg:amcmc_modified}. In particular, in Theorem \ref{theorem ARGS convergence} we have proved ergodicity of the modified ARSGS Algorithm \ref{alg:ARSGS ergodic}.

In order to establish convergence in Theorem \ref{theorem ARGS convergence}, we do not require the sequence of estimated optimal sampling probabilities weights $p^n$ to converge at all. Instead, we require only the diminishing condition to hold, i.e., $|p_n - p_{n-1}| \overset{P}{\rightarrow} 0$ as $n\to \infty$. In fact,  it is not clear whether the estimated weights converge to the pseudo-optimal ones, even if one knows the target covariance matrix. Empirically, for numerous examples, we have observed that the adapted selection probabilities do converge to a unique solution, where the uniqueness is guaranteed by Theorem \ref{theorem uniqueness}.

{\bf Open problem. } Assume that the covariance matrix of the target distribution is known, i.e., $\Sn = \Sigma$ for all $n$. Prove that the estimated weights $p^i$ in the ARSGS algorithm converge to the pseudo-optimal weights $\po$.

We emphasise that there is no universal algorithm to optimise the pseudo-spectral gap function (\ref{target function}), given that the covariance structure $\Sigma$ is unknown.

Various other modification of the ARSGS algorithm are possible. 
For instance, we can think of using some other optimisation algorithm instead of the subgradient method (described in Algorithm \ref{alg:subgradient method})  in order to estimate  the pseudo-optimal weights (\ref{definition pseudo-optimal}). Also, the user may know the structure of the covariance matrix $\Sigma$ in advance, so that the naive estimator (\ref{naive estimator}) could be improved. For example, if the covariance matrix is banded, a more efficient threshold estimator should be used (see \cite{Bickel2008}). 

In Section 5.1 of \cite{Chimisov2018} we introduce a modified (Air) version of the ARSGS Algorithm \ref{alg:ARSGS ergodic}, for which we prove the SLLN and the Mean Squared Error convergence under the local simultaneous geometric drift assumption \ref{cond:local_simultaneous_drift}. If, additionally, the sequence of adapted selection probabilities $p^n$ converges, we derive the CLT.

\section{SUPPLEMENTARY  MATERIAL}\label{SUPPLEMENTARY  MATERIAL}
\hfill \break
\noindent {\bf Proof of Lemma \ref{lemma hermite polynomials}.} The proof is a modification of Theorem 1 of \cite{Amit1996}, and thus we outline only the key points.

One can easily check that $\{H_\alpha (Kx)\}$ form an orthonormal system in $L_2 (\mathbb{R}^d,\pi)$ using the definition. From Theorem 6.5.3 of \cite{Roy1999}, it follows that one dimensional Hermite polynomials $h_k$ form a complete orthogonal basis of $L_2(\mathbb{R}, \exp(-x^2/2))$, implying  $\{H_\alpha (Kx)\}$ form an orthogonal basis of $L_2 (\mathbb{R}^d,\pi)$.  

For $c\in \mathbb{R}^d$, define a generating function
\begin{align}\label{generating_function}
f_c (x) :=\sum_{\alpha}c^{\alpha} \frac{H_\alpha(Kx)}{\sqrt{\alpha!}},
\end{align}
where $c^\alpha:= c_1^{\alpha_1},.., c_d^{\alpha_d}$ and $0^0 :=1$.

From Section 4.2.1 of \cite{Steven1984}, we know that the generating function can be represented as
$$f_c(x) = \exp \( \<c,K x\>- \frac{\| c \|^2}{2} \).$$

Recall that $Pr_i$ stands for full conditional update of $x_i = (x_{i1},.. x_{ir_i})$ from its full conditional. For functions $f\in L_2 (\mathbb{R}^d,\pi)$, let
$$(Pr_i f)(x):= \int f(y_i,x_{-i}) \pi(y_i| x_{-i}) d y_i.$$

Define $T^{(i)} := I-K D_i   K$. Note that $T^{(i)}  = \(T^{(i)}_{ij}\)_{i,j =1}^d$ is a $d-$dimensional matrix.

The key property to prove the first part of the lemma is the following statement that can be obtained via direct calculations.

\begin{lemma} \label{lemma f_c}
	For all $c\in \mathbb{R}^d$
	$$(Pr_i f_c) (x)=f_{T^{(i)}c} (x),$$
	where $f_{c}$ is defined in (\ref{generating_function}).
\end{lemma}

Let $\Pi_k$ be the set of all sequences $(\epsilon_1,..,\epsilon_k)$ of length $k$ with elements from $\{1,..,d\}$. Partition $\Pi_k$ into equivalence classes $R_\alpha$, $\alpha = (\alpha_1,..,\alpha_d) \in Z^n_+$, $|\alpha|=k$, such that $\epsilon = (\epsilon_1,..,\epsilon_k) \in R_\alpha$ if and only if the sequence $\epsilon$ has $\alpha_1$ 1's,.., $\alpha_d$ d's. In other words, $R_\alpha$ forms a set of all permutations of the elements of $(\epsilon_1,..,\epsilon_k)$, implying that the number of elements in $R_\alpha$  is $|R_\alpha| = \frac{k!}{\alpha!}$.

Lemma \ref{lemma f_c} implies

\begin{align} \label{image 1}
Pr_i \(\sum_{\alpha}c^{\alpha} \frac{H_\alpha(K \cdot)}{\sqrt{\alpha!}}\)(x)=\sum_{\alpha} (T^{(i)}c)^\alpha\frac{H_\alpha(Kx)}{\sqrt{\alpha!}}.
\end{align}
Fix $\beta$ such that $|\beta|=k$. For each $\alpha$, $|\alpha|=k$, fix some representative  $\sigma = \sigma (\alpha)\in R_\alpha$. Rewrite $T^{(i)}c$ as
$$T^{(i)}c = \(T^{(i)}_{j1}c_1+..+T^{(i)}_{jd} c_d\)_{j=1}^d.$$
Since $c\in \mathbb{R}^d$ is arbitrary, the coefficient of $c^{\beta}$ on both sides of (\ref{image 1}) should coincide for all $\beta \in Z^d_+$, providing a formula for the image of $H_\beta (Kx)$:

\begin{align*}
Pr_i \(\frac{H_\beta (K \cdot)}{\sqrt{\beta!}}\)(x) =\sum_{|\alpha| = k} \frac{1}{\sqrt{\alpha!}} \( \sum_{\epsilon \in R_\beta} T^{(i)}_{\sigma_1 \epsilon_1}.. T^{(i)}_{\sigma_k \epsilon_k}\) H_\alpha(Kx).
\end{align*}
Since $\sigma = \sigma(\alpha)$ was chosen arbitrary, the above sum is equal to

\begin{align*}
&Pr_i \(\frac{H_\beta (K \cdot)}{\sqrt{\beta!}}\)(x) = \sum_{|\alpha| = k} \frac{1}{|R_\alpha| \sqrt{\alpha!}}\(\sum_{\epsilon\in R_\beta,\ \sigma\in R_\alpha} T^{(i)}_{\sigma_1 \epsilon_1}.. T^{(i)}_{\sigma_k \epsilon_k}\) H_\alpha(Kx) =\\
&= \sum_{|\alpha| = k} \frac{\sqrt{\alpha !}}{k!}\(\sum_{\epsilon\in R_\beta,\ \sigma\in R_\alpha} T^{(i)}_{\sigma_1 \epsilon_1}.. T^{(i)}_{\sigma_k \epsilon_k}\) H_\alpha(Kx).
\end{align*} 
We conclude that  
\begin{align}\label{image of L}
Pr_i (H_\beta (K \cdot))(x) =\sum_{|\alpha| = k} \frac{\sqrt{\alpha!} \sqrt{\beta!}}{k!}\(\sum_{\epsilon\in R_\beta,\ \sigma\in R_\alpha} T^{(i)}_{\sigma_1 \epsilon_1}.. T^{(i)}_{\sigma_k \epsilon_k}\) H_\alpha (K x),
\end{align}
implying the first part of the Lemma.
\\

We are left to  show that the maximum eigenvalue of $P_p$ on $S_k$ is non-increasing, revealing that the second largest eigenvalue of $P_p$ is attained on $S_1$. 

Recall that $\{R_\alpha|\ \alpha\in Z^n_+,\ |\alpha|=k\}$ form a partition of all possible sequences $(\epsilon_1,..,\epsilon_k)$, $\epsilon_i\in \{1,..,d\}$. Moreover,  as we have just seen, $Pr_i$ is invariant on $S_k$ ($S_k$ is defined in the statement of the lemma), that is $Pr_i$ acts like a matrix on $S_k$. (\ref{image of L}) implies that $Pr_i$ can be represented as

$$Pr_i (H_\beta (K \cdot))(x) =\sum_{|\alpha| = k} \frac{1}{\sqrt{|R_\alpha| |R_\beta|}}\(\sum_{\epsilon\in R_\beta,\ \sigma\in R_\alpha} T^{(i)}_{\sigma_1 \epsilon_1}.. T^{(i)}_{\sigma_k \epsilon_k}\) H_\alpha (K x),$$

implying that the matrix that corresponds to $Pr_i$ consists of entries 
$$\frac{1}{\sqrt{|R_\alpha| |R_\beta|}}\(\sum_{\epsilon\in R_\beta,\ \sigma\in R_\alpha} T^{(i)}_{\sigma_1 \epsilon_1}.. T^{(i)}_{\sigma_k \epsilon_k}\)$$
Thus we have shown that on $S_k$, $P_p$ acts as a matrix with corresponding entries obtained as normalised block sums of 
$$F_k=\sum_{i=1}^s p_i\ (T^{(i)})^{\otimes k},$$
where $(T^{(i)})^{\otimes k}$ is the $k-th$ Kronocker product of $T^{(i)}$ (i.e., the $k-th$ tensor product, see \cite{Simon1980}, VIII.10).

The next statement is Lemma 1 from \cite{Amit1996} and we do not prove it.
\begin{lemma}\label{lemma normilized sums}
	Let $A$ be a non-negative definite $r\times r$ matrix. Let $R_1,.., R_q$ be a partition of $\{1,..,r\}$. Define matrix $B$ to be the $q\times q$ matrix, 
	$$B_{kl} = \frac{1}{\sqrt{|R_k| |R_l|}}  \sum_{i\in R_k,\ j\in R_l} A_{ij}.$$
	Then the maximum eigenvalue of $B$ is less or equal than the maximum eigenvalue of $A$.
\end{lemma}

Lemma \ref{lemma normilized sums} shows that the maximum eigenvalue of $P_p$ restricted to $S_k$ is dominated by the maximum eigenvalue of $F_k$.

Rewrite $F_{k+1}$ as a difference of two positive semi-definite operators 
$$F_{k+1} = F_k\otimes I - \sum_{i=1}^s p_i \(T^{(i)}\)^{\otimes k}\otimes \(I - T^{(i)}\)$$

It follows  that for all $k\geq 0$, $F_k\otimes I \geq F_{k+1}$  (i.e., for all vectors $x$, $\<F_k\otimes I x, x\>\geq \<F_{k+1}x, x\>$). Since $\|F_k\| = \|F_k \otimes I\|$ (see \cite{Simon1980}, VIII.10), the largest eigenvalue of $F_k$ (that is equal to largest eigenvalue of $F_k\otimes I$) is greater than the one of the operator $F_{k+1}$. Thus the largest eigenvalues of $P_p|_{S_k}$, $k\geq 0$ form a non-increasing sequence.

Note that $P_p|_{S_0}$ corresponds to the unit eigenvalue and that the matrix that corresponds to $P_p|_{S_1}$ is exactly $F_1$, as easily seen from (\ref{image of L}). Therefore, the second largest eigenvalue of $P_p$ is attained on $S_1$ and is equal to the maximum eigenvalue of $F_1$. \\
$\square$\\

\noindent {\bf Proof of Theorem \ref{theorem spectral gap for normal distribution}.} From the formula (\ref{image of L}) it follows that for $k = 1$, a matrix that corresponds to  $P_p|S_1$ is exactly $F_1$.\\
$\square$\\

\noindent {\bf Proof of Lemma \ref{lemma isometry argument}.}
Let $\lambda\neq 0$ such that
$$AB x = \lambda x$$ 
for some non-zero $x$. Multiply both sides by $B$ 
$$BA (Bx) = \lambda B x.$$
If $\lambda = 0$ and $x \neq 0$ such that
$$AB x = 0,$$
we may have either $Bx \neq 0$ or $Bx = 0$. In the first case multiply both sides by $B$ so that $BA (Bx) = 0$. Otherwise, if $A$ is invertible, find $y$ such that $Ay = x$ so that $BA y =0$. If $A$ is not invertible, there exists $y \neq 0$ such that $Ay=0$ so that again $BAy=0$.\\
$\square$\\

\noindent { \bf Proof of Theorem \ref{theorem uniqueness}.} Define

\begin{align} \label{theorem uniqueness:target function}
h(p)=\lm \(D_p Q \)=\lm \(\sqrt{Q}  D_p    \sqrt{Q} \).
\end{align}
Assume $\po_1$ and $\po_2$ are two different points that maximise $h$. Then a function 
$$g(\alpha) := h(\alpha \po_1 + (1-\alpha)\po_2)$$
is constant on $[0,1]$ due to concavity of $h$ (see  Proposition \ref{proposition concavity}) and equals, say, $\lambda$. Since $h(p)$ is itself the minimum eigenvalue of $\sqrt{Q}D_p\sqrt{Q}$, there exist unit vectors $x_0$, $x_1$, $x_2$ such that
\begin{align*}
&\sqrt{Q}D_{\frac{1}{2} \po_1 + \frac{1}{2}\po_2}\sqrt{Q} x_0 = \lambda x_0\\
&\sqrt{Q}D_{\po_1}\sqrt{Q} x_1 = \lambda x_1\\
&\sqrt{Q}D_{\po_2}\sqrt{Q} x_2 = \lambda x_2
\end{align*}
Since $g$ is constant on $[0,1]$,
\begin{align*}
& \frac{1}{2} \<\sqrt{Q}D_{\po_1}\sqrt{Q} x_0, x_0\> + \frac{1}{2} \<\sqrt{Q}D_{\po_2}\sqrt{Q} x_0, x_0\> = g\(\frac{1}{2}\) = \frac{g(0)}{2} + \frac{g(1)}{2} = \\
&=\frac{1}{2} \<\sqrt{Q}D_{\po_1}\sqrt{Q} x_1, x_1\> + \frac{1}{2} \<\sqrt{Q}D_{\po_2}\sqrt{Q} x_2, x_2\>\leq \\
&\leq \frac{1}{2} \<\sqrt{Q}D_{\po_1}\sqrt{Q} x_0, x_0\> + \frac{1}{2} \<\sqrt{Q}D_{\po_2}\sqrt{Q} x_0, x_0\>,
\end{align*}
where the last inequality holds since $x_1$, $x_2$ are the minimum eigenvectors. Hence, 
\begin{align*}
& \<\sqrt{Q}D_{\po_1}\sqrt{Q} x_0, x_0\> = \<\sqrt{Q}D_{\po_1}\sqrt{Q} x_1, x_1\> = \lambda,  \\
&  \<\sqrt{Q}D_{\po_2}\sqrt{Q} x_0, x_0\> =  \<\sqrt{Q}D_{\po_2}\sqrt{Q} x_2, x_2\> = \lambda.
\end{align*}
Therefore,
\begin{align} \label{same vector pseudo-points}
& \sqrt{Q}D_{\po_1}\sqrt{Q} x_0 = \lambda x_0 \mbox{ and } \sqrt{Q}D_{\po_2}\sqrt{Q} x_0 = \lambda x_0,
\end{align}
which follows from the following simple statement.

\begin{lemma}\label{lemma min eigenvector} Let $A$ be a $n\times n$ symmetric matrix, and $x$ be a unit vector, such that 
	$\<Ax, x\> = \lm (A)$. Then $Ax = \lm (A) x$.
\end{lemma}
Let $y_0 = \sqrt{Q} x_0$. Then (\ref{same vector pseudo-points}) is equivalent to

\begin{align*} 
& D_{\po_1} y_0 = \lambda \Sigma y_0 \mbox{ and }  D_{\po_2} y_0 = \lambda \Sigma y_0.
\end{align*}
Using the definition of (\ref{matrix_D_p}), the last equalities yield
$$(\po_i)_j = \lambda \frac{\<\(\Sigma y_0\)_j, (y_0)_j\>}{\<Q^{-1}_{jj} (y_0)_j, (y_0)_j\>},$$
for $i=1, 2$, if $ (y_0)_j\neq 0$. 

Let $p:=\frac{1}{2}\po_1 + \frac{1}{2} \po_2$. It is left to show that for every $j\in\{1,..,s\}$, one can find a minimum eigenvector $x_0$ of $\sqrt{Q} D_p \sqrt{Q}$, such that for the corresponding vector $y_0 = \sqrt{Q} x_0$, we have  $(y_0)_j \neq 0$.

Define a  space $S_0 = \{x_0 | \sqrt{Q} D_p \sqrt{Q} x_0 = \lambda x_0\}$ as a space generated by all the minimum eigenvectors of $\sqrt{Q} D_p \sqrt{Q}$. 

Assume, on the contrary, that for some $j\in \{1,.., s\}$,  and all $x_0\in S_0$, we have $ (y_0)_j = 0$.

Define a space $S_\perp:=\{ x \ | x \mbox{ is orthogonal to } S_0 \}$. For  $\epsilon \geq 0$, let 
$$A_\epsilon :=\sqrt{Q}  D_{p^{(\epsilon)}} \sqrt{Q},$$
where $p^{(\epsilon)}_k =  (1+\epsilon) p_k $ if $k\neq j$, and $p^{(\epsilon)}_j = p_j - \epsilon\sum_{k\neq j} p_k$. Note that for all $x_0 \in S_0$ and $\epsilon\geq 0$,
\begin{align}\label{minimum eigenvalue}
A_\epsilon x_0 = \sqrt{Q} D_{p^{(\epsilon)}} \sqrt{Q}x_0 = \sqrt{Q} D_{p^{(\epsilon)}}y_0 = (1+\epsilon)\lambda \sqrt{\Sigma}y_0 = (1+\epsilon)\lambda x_0,
\end{align}
since $y_0 = \sqrt{Q} x_0$, $D_{p} y_0 = \lambda \Sigma y_0$, and $(y_0)_j=0$ by the assumption. That is, $(1+\epsilon)\lambda$ is an eigenvalue of $A_\epsilon$, and $S_0$ is the subspace of the corresponding eigenvectors.

Also, $S_\perp$ is invariant under $A_\epsilon$ (i.e., $A_\epsilon S_\perp \subset S_\perp$). Indeed, for all $x \in S_\perp$ and $x_0 \in S_0$,

$$\< A_\epsilon x, x_0\> = \<  x, A_\epsilon x_0\> = \lambda (1+\epsilon) \<x, x_0\> = 0.$$

Let ${\lambda}^{(\epsilon)}$ be the minimum eigenvalue of $A_\epsilon$ restricted to the space $S_\perp$. Since $S_0$ contains all possible minimum eigenvectors of $\sqrt{Q} D_{p} \sqrt{Q}$, we have ${\lambda}^{(0)}  > \lambda$. Therefore, we can find small enough $r>0$, such that 
$${\lambda}^{(0)} > (1+r)\lambda.$$

Recall that ${\lambda}^{(\epsilon)}$ is a continuous function of $\epsilon$ (since it is a concave function by the Proposition \ref{proposition concavity}). Thus there exists $\delta \in (0, r)$, such that for all $\epsilon\in [0,\delta]$, ${\lambda}^{(\epsilon)} >(1+r)\lambda$ and also $p^{(\epsilon)} \in \Delta_{s-1}$. In particular, 
$${\lambda}^{(\delta)} >(1+r)\lambda>(1+\delta)\lambda.$$

Since $S_\perp \cup S_0 = \mathbb{R}^d$, and $A_\delta$ is a symmetric, invariant operator on $S_\perp$ and $S_0$, we obtain,

$$\lm(A_\delta) = \min \{\lm(A_\delta |_{S_\perp}), \lm(A_\delta |_{S_0})\} =  \min \{{\lambda}^{(\delta)}, (1+\delta)\lambda\} = (1+\delta)\lambda, $$
meaning   $\lambda (1+\delta)$ is the minimum eigenvalue of $A_\delta = \sqrt{Q}  D_{p^{(\delta)}} \sqrt{Q}$, $p^{(\delta)}\in \Delta_{s-1}$. Hence $\lambda$ is not the maximum of (\ref{theorem uniqueness:target function}), which contradicts to the definition of $\lambda$.  Thus there exists $x_0 \in S_0$, such that $(y_0)_j \neq 0.$
\\
$\square$\\

\noindent  {\bf Proof of Lemma \ref{lemma min eigenvector}.}  Let $\{x_i\}$ be an orthonormal basis of eigenvectors of $A$ with $\{\lambda_i\}$ being the corresponding eigenvalues. Then $x = \sum_{i=1}^n \<x,x_i\>x_i$. We need to show that $\<x,x_i\> = 0$ for all $x_i$ that are not the minimum eigenvectors. Assume there are at least two vectors $x_{k}$ and $x_{j}$ such that $\lambda_k \neq \lambda_j$, $\<x,x_k\>^2>0$, and  $\<x,x_j\>^2>0$. Then
$$\<Ax, x\> = \sum_{i=1}^n \lambda_i \<x, x_i\>^2> \lm(A) \sum_{i=1}^n \<x, x_i\>^2 = \lm(A),$$
contradicting the assumption that $\<Ax, x\> = \lm(A)$.\\
$\square$\\

\noindent { \bf Proof of Theorem \ref{theorem upper bound}.} Let $P_p$ be the Gibbs kernel as in (\ref{kernel}) with corresponding weights $p$ and let $P_{\frac{1}{s}}$ be the kernel of the vanilla chain. For functions $f, g \in L_2(\mathbb{R}^d, \pi)$ let 
$$<f, g> = \int f g {\rm d} \pi,\ \|f\|^2 = \int f^2 {\rm d} \pi.$$ 
Using an equivalent representation for the spectral gap (see a remark to Theorem 2 of \cite{Roberts1997b}), inequality (\ref{gap upper bound}) is equivalent to
\begin{align*}
\inf_{\|f\|=1, \pi(f) = 0}\Bigg<(I-P_p)f, f\Bigg> \leq \max_i \(\frac{p_i}{q_i}\) \inf_{\|f\|=1, \pi(f) = 0}\Bigg<(I-P_{q})f, f\Bigg> 
\end{align*}  
It suffices to establish
$$  \Bigg<(I-P_p)f, f\Bigg>\leq \max_i \(\frac{p_i}{q_i}\)  \Bigg<(I-P_{q})f, f\Bigg> $$
for all $f$, $\|f\|=1, \pi(f) = 0$. Let $j =\underset{i}{\argmax}\ \frac{p_i}{q_i}$ Using the representation (\ref{kernel}) of $P_p$, the last inequality is equivalent to
$$ \sum_{i=1}^s p_i\(\frac{q_i }{p_i} - \frac{q_j }{p_j} \)\<Pr_i f, f\>+ \frac{q_j}{ p_j}\|f\|^2 \leq \|f\|^2.$$
Since $\frac{q_j }{p_j}\leq \frac{q_i }{p_i}$ and $\<Pr_i f, f\> \leq \|f\|^2$, the last inequality follows:

$$\sum_{i=1}^s p_i\(\frac{q_i }{p_i} - \frac{q_j }{p_j} \)\<Pr_i f, f\>+ \frac{q_j}{ p_j}\|f\|^2\leq \sum_{i=1}^s p_i\(\frac{q_i }{p_i} - \frac{q_j }{p_j} \) \|f\|^2 + \frac{q_j}{ p_j}\|f\|^2 =  \|f\|^2,$$
where in the last equality we used the fact that $\sum_{i=1}^s p_i= \sum_{i=1}^s q_i = 1$.\\
$\square$\\

\noindent  { \bf Proof of Proposition \ref{proposition example 2k*2k}.}  For $i=1,..,k$ let

$$A_i = \left( \begin{array}{cc} p_{2i-1}  & p_{2i-1} \rho_i  \\
p_{2i} \rho_i & p_{2i} \\
\end{array} \right).$$

One can see that the pseudo-optimal weights $\po$ satisfy
\begin{align}\label{optimal k}
\po = \argmax_{p\in \Delta_{2k-1}} \min\{\lm (A_1-\lambda I), ... , \lambda_{\min} (A_k-\lambda I)\}.
\end{align}

Set $\alpha_i = p_{2i-1}+p_{2i}$, $i=1,..,k$. We obtain

$$\argmax_{p\in \Delta_{2k-1}} \lambda_{min} (A_i-\lambda I)=\po_{2i}=\po_{2i-1}=\frac{\alpha_i}{2},$$
so that (\ref{optimal k}) takes the form
\begin{align}\label{optimal k 2}
\po = \argmax_{p\in \Delta_{2k-1}} \min\left \{\frac{\alpha_1 (1-\rho_1)}{2}, ... ,\frac{\alpha_k (1-\rho_k)}{2}\right\}.
\end{align}
It is easy to verify that $\po$ should satisfy
$$\frac{\alpha_1 (1-\rho_1)}{2}=...=\frac{\alpha_k (1-\rho_k)}{2}.$$
The last relation leads to (\ref{alpha k}) and we conclude that the optimal selection probabilities $\po_i$, $i=1,..,k$ are computed as in (\ref{expression optimal k}). Finally, (\ref{optimal rate of convergence example}) follows from (\ref{optimal k 2}).
\\
$\square$\\

\noindent  {\bf Proof of Proposition \ref{proposition extended pseudo-optimal points}.} Note that the target function $f$ in (\ref{target function}) is the minimum eigenvalue of 

$$ \left( \begin{array}{cc} \sqrt{Q} D_w \sqrt{Q} & 0  \\
0 & 1 - w_1-..-w_s
\end{array} \right),$$
so that we can rewrite $f$ as
\begin{equation}\label{sp_gap:block_structure}
\begin{split}
&f(w)=\lm\(\Dt \Qt\) = \lm \(\sqrt{\Qt} \Dt \sqrt{\Qt}\) =\\
&= \min\{\lm(\sqrt{Q} D_w \sqrt{Q}), 1 - w_1-..-w_s\} = \\
& = \min\{\PG(w), 1 - w_1-..-w_s\}.
\end{split}
\end{equation}
Let $w^{\star} = \underset{{w\in \Delta_{s}}}\argmax f(w)$, and denote $p^{\star}_j = \frac{w^{\star}_j}{w^{\star}_1+.. + w^{\star}_s},\ j=1,..,s,$. To prove the proposition it suffices to show that for the pseudo-optimal weights $\po$, 
\begin{align}\label{sp_gap:block_structure:ineq}
\PG (\po)\leq \PG (p^{\star}).
\end{align}
Let $k^{\star} = \sum_{i=1}^s w^{\star}_i$. It is easy to see from (\ref{sp_gap:block_structure}) that 
$$1 - k^{\star} = \PG (w^{\star}) = \PG(k^{\star} p^{\star}) = f(k^{\star} p^{\star}).$$

Since for any $k>0$ and any $p\in\Delta_{s}$, $\PG(k p) = k \PG(p)$, we can choose $k<1$ such that 
$$1 - k = \PG(k\po) = f(k\po).$$
Hence, by definition of $w^{\star}$,
$$\PG(k \po)  = 1 - k \leq  1 - k* = \PG(k^{\star} p^{\star}),$$
implying $k^{\star} \leq k$. Therefore,
\begin{align*}
&\PG(\po) = \frac{1}{k} \PG(k \po) \leq \frac{1}{k} \PG(k^{\star} p^{\star}) = \\
& = \frac{k^{\star}}{k} \PG( p^{\star}) \leq  \PG( p^{\star}) ,
\end{align*}
whence we conclude (\ref{sp_gap:block_structure:ineq}).\\
$\square$\\

\noindent  {\bf Proof of Proposition \ref{proposition concavity}.} Note that $\<\sqrt{\Qnt} \Dnt \sqrt{\Qnt} x, x\>$ is linear in $w$ for all $x\in \mathbb{R}^{d+1}$. That is, there exist functions $a_0(x),.., a_{s}(x)$ such that 
$$\<\sqrt{\Qnt} \Dnt \sqrt{\Qnt} x, x\>=a_0(x)+w_1 a_1(x)+..+w_{s} a_{s}(x).$$

Thus $\<\sqrt{\Qnt} \Dnt \sqrt{\Qnt} x, x\>$ is concave for all $x$. Then $\f$ is concave as the minimum over concave functions.\\
$\square$\\

\noindent  {\bf Proof of Proposition \ref{proposition ergodicity MSM}.} Geometric ergodicity follows if we find drift coefficients to establish \ref{cond:geometric_drift}. We argue that 

$$V(x_{1:n}) = \frac{x_1^2}{2}+ x_2^2..+x_{n-1}^2+\frac{x_n^2}{2}$$
is an appropriate drift function for the vanilla RSGS is cases \ref{msm_a}, \ref{msm_b} and \ref{msm_c}. Note that $V$ does not depend on the regimes $r(i)$. One can work out the full conditionals for $X_i$, 

\begin{equation}\label{msm:full_cond_state}
\begin{split}
&X_i | X_{-i}, Y_{1:n}, r(1),..,r(n) \sim N(\mu, q_{r(i),r(i+1)}),\\
&\mu  = q_{r(i),r(i+1)}\(\frac{X_{i-1}I_{\{i>1\}}}{\sigma^2_{r(i)}}+\frac{X_{i+1}I_{\{i<n\}}}{\sigma^2_{r(i+1)}}+\frac{Y_{i}}{\beta^2}\),
\end{split}
\end{equation}
where 
$$q_{r(i),r(i+1)} = \frac{1}{\frac{1}{\beta^2}+\frac{I_{\{i>1\}}}{\sigma^2_{r(i)}}+\frac{I_{\{i<n\}}}{\sigma^2_{r(i+1)}}}.$$
Here we set $X_{n+1} = X_{0}:= 0.$

Let $f_i(x_{1:n}) = x_i^2$. For $Pr_i$, $i = 1,.., n$, defined in (\ref{kernel}), where $Pr_i$ corresponds for updating $X_i$ from its full conditional, one see it is obvious that
\begin{align}\label{msm:drift_1}
Pr_j f_i(x_{1:n}) =  f_i(x_{1:n}),\ i\neq j.
\end{align}
 From (\ref{msm:full_cond_state}), using the Cauchy-Schwartz inequality, we get
\begin{equation}\label{msm:drift_2}
\begin{split}
&Pr_i f_i (x_{1:n}) = \mu^2  + q_{r(i),r(i+1)} \leq \\
&\leq q^2_{r(i),r(i+1)} \(\frac{I_{\{i>1\}}}{\sigma^4_{r(i)}}+\frac{I_{\{i<n\}}}{\sigma^4_{r(i+1)}}\) \times\\
&\times \Big(f_{i-1}(x_{1:n}) I_{\{i>1\}} + f_{i+1}(x_{1:n}) I_{\{i<n\}} \Big) + L_i(x_{i-1},x_{i+1}),
\end{split}
\end{equation}
where $L_i(x_{i-1},x_{i+1})$ is a linear function. Note that for the considered cases \ref{msm_a}, \ref{msm_b} and \ref{msm_c}, for any configuration of $(r(1),.., r(n))$ and $i \in \{2,..,n-1\}$, 
\begin{align}\label{msm:drift_coeff_bound_1}
2 q^2_{r(i),r(i+1)} \(\frac{1}{\sigma^4_{r(i)}} + \frac{1}{\sigma^4_{r(i+1)}}\)<0.99.
\end{align}
Moreover, for $i\in\{1 , n\}$,
\begin{align}\label{msm:drift_coeff_bound_2}
q^2_{r(i),r(i+1)} \(\frac{I_{\{i>1\}}}{\sigma^4_{r(i)}}+\frac{I_{\{i<n\}}}{\sigma^4_{r(i+1)}}\) \leq 0.57.
\end{align}
It follows from (\ref{msm:drift_2}), (\ref{msm:drift_coeff_bound_1}) and (\ref{msm:drift_coeff_bound_2}), 

\begin{equation}\label{msm:drift_3}
\begin{split}
& \frac{1}{2}Pr_1 f_1 (x_{1:n}) + \sum_{i = 2}^{n-1} Pr_i f_i (x_{1:n}) + \frac{1}{2} Pr_n f_n (x_{1:n}) \leq \\
& \leq  0.99 V(x_{1:n}) + L(x_{1:n}),
\end{split}
\end{equation}
where $L(x_{1:n})$ is a linear function. Together with (\ref{msm:drift_1}), the inequality (\ref{msm:drift_3}) yields

\begin{align}\label{msm:drift_4}
\frac{1}{d} \sum_{i=1}^n Pr_i V(x_{1:n}) \leq \frac{\lambda}{2} V(x_{1:n}) + A,
\end{align}
for some $\lambda <1$ and $A<\infty$.

For $Pr_{i + n}$ that corresponds for updating $r(i)$ from its full conditional, since $V$ does not depend on $r(i)$, we get
\begin{align}\label{msm:drift_5}
Pr_{i+n} V  = V.
\end{align}

Let $P_{\frac{1}{d}}$ be the RSGS kernel that corresponds to the vanilla chain with uniform sampling weights $\frac{1}{d}$. Combining (\ref{msm:drift_4}) and (\ref{msm:drift_5}) together, we obtain, 

$$P_{\frac{1}{d}} V \leq \frac{\lambda}{2} V + \frac{1}{2} V + A.$$
Since for all $C<\infty$, set $\{\(x_{1:n},r(1),..,r(n)\)\ |\ V(x_{1:n})<C\}$ is small, Lemma 15.2.8 of \cite{Meyn2009} yields that $V$ is a geometric drift function.
\\

For the RSGS with non-uniform selection probabilities $p = (p_1,.. , p_d)$, we can use theorem \ref{theorem local simultaneous drift} to conclude the geometric ergodicity.\\
$\square$\\

\end{document}